\documentclass[aps,prd,preprint,a4paper,showpacs,nofootinbib,superscriptaddress]{revtex4-2}
\usepackage{bm}
\usepackage{indentfirst}
\usepackage{amsmath}
\usepackage{graphicx}
\usepackage{float}
\usepackage{amssymb}
\usepackage{subfigure}
\usepackage{amssymb}
\usepackage{hyperref}
\usepackage{epstopdf}
\usepackage[section]{placeins}

\usepackage[utf8]{inputenc}
\hypersetup{
    colorlinks=true,
    linkcolor=red,
    citecolor=blue,
}
\usepackage{color}
\usepackage[T1]{fontenc}
\usepackage{txfonts}
\usepackage{orcidlink}

\begin{document}

\baselineskip=0.5 cm

\title{Detection of scalar fields by Extreme Mass Ratio Inspirals with Kerr black hole }
\author{Hong Guo}
\email{gh710105@gmail.com}
\address{\textit{Center for Gravitation and Cosmology, College of Physical
Science and Technology, Yangzhou University, Yangzhou 225009, China}}
\address{\textit{Shanghai Frontier Research Center for Gravitational Wave Detection, Shanghai Jiao Tong
University, Shanghai 200240, China}}

\author{Yunqi Liu}
\email{yunqiliu@yzu.edu.cn (corresponding author)}
\address{\textit{Center for Gravitation and Cosmology, College of Physical
Science and Technology, Yangzhou University, Yangzhou 225009, China}}
\address{\textit{School of Aeronautics and Astronautics, Shanghai Jiao Tong
University, Shanghai 200240, China}}

\author{Chao Zhang}
\email{chao_zhang@hust.edu.cn}
\address{\textit{School of Physics, Huazhong University of Science and Technology, Wuhan, Hubei 430074, China}}

\author{Yungui Gong\,\orcidlink{0000-0001-5065-2259}}
\email{yggong@hust.edu.cn}
\address{\textit{School of Physics, Huazhong University of Science and Technology, Wuhan, Hubei 430074, China}}

\author{Wei-Liang Qian}
\email{wlqian@usp.br}
\address{\textit{Escola de Engenharia de Lorena, Universidade de S\~ao Paulo, 12602-810, Lorena, SP, Brazil}}
\address{\textit{Faculdade de Engenharia de Guaratinguet\'a, Universidade Estadual Paulista, 12516-410, Guaratinguet\'a, SP, Brazil}}
\address{\textit{Center for Gravitation and Cosmology, College of Physical
Science and Technology, Yangzhou University, Yangzhou 225009, China}}

\author{Rui-Hong Yue}
\email{rhyue@yzu.edu.cn}
\address{\textit{Center for Gravitation and Cosmology, College of Physical
Science and Technology, Yangzhou University, Yangzhou 225009, China}}

\begin{abstract}
We study extreme mass ratio inspirals occurring in modified gravity, for which the system is modeled by a small compact object with scalar charge spiraling into a supermassive Kerr black hole.
Besides the tensorial gravitational waves arising from the metric perturbations, radiations are also induced by the scalar field.
The relevant metric and scalar perturbations are triggered by the orbital motion of the small object, which give rise to a system of inhomogeneous differential equations under the adiabatic approximation.
Such a system of equations is then solved numerically using Green's function furnished by the solutions of the corresponding homogeneous equations.
To explore the present scenario from an observational perspective, we investigate how the pertinent observables are dependent on specific spacetime configurations.
In this regard, the energy fluxes and the gravitational wave dephasing accumulated during the process are evaluated, as functions of the scalar charge, mass ratio, and spin of the central supermassive black hole.
In particular, the presence of additional scalar emission leads to a more significant rate of overall energy loss, which, in turn, decreases the total number of orbital cycles before the small object plunges into the central black hole.
Moreover, for a central black hole with a higher spin, the imprints of the scalar charge on the resultant gravitational radiations are found to be more significant, which indicates the possibility of detecting the scalar charge.
\end{abstract}

\maketitle

\section{Introduction}

Since the first gravitational wave (GW) detection in 2015 \cite{LIGOScientific:2016aoc}, the Laser Interferometer Gravitational Wave Observatory (LIGO) Scientific Collaboration \cite{Harry:2010zz,TheLIGOScientific:2014jea}, the Virgo Collaboration \cite{TheVirgo:2014hva}
and the Kamioka Gravitational Wave Detector (KAGRA) Collaboration \cite{Somiya:2011np,Aso:2013eba} have detected tens of coalescing events of binary systems consisting of binary black holes, binary neutron stars,
and neutron-star black hole (BH) binaries \cite{LIGOScientific:2018mvr,Abbott:2020niy,LIGOScientific:2021djp}.
GW detections provide a new window to measure the properties of compact binaries and test new theories of gravity in the strong-field and nonlinear regimes \cite{Barausse:2016eii, Babak:2017tow, Sopuerta:2009iy, Pani:2011xj, Yunes:2011aa, Yagi:2016jml, Berti:2018cxi, Barack:2018yly, Barack:2006pq, Pani:2010em}.
The success of the ground-based GW projects has raised further confidence and expectation in space-based GW detectors such as the Laser Interferometer Space Antenna (LISA) \cite{LISA:2017pwj,Team_1997}, TianQin \cite{TianQin:2015yph}, Taiji \cite{Hu:2017mde}, and DECIGO \cite{Kawamura:2006up}. 
A recent review on the concepts and status of space-borne gravitational wave detection can be found in \cite{Gong:2021gvw}. 

Extreme mass ratio inspirals (EMRIs) are among the most promising sources for space-based GW detectors \cite{Gair:2012nm}.
An EMRI occurs when a stellar-mass object (the secondary) is trapped in a sufficiently tight orbit of a supermassive BH (the primary).
Due to gravitational radiation, the orbital energy decays, and subsequently, the secondary body will spiral until it eventually plunges into the primary BH.
Typically, the timescale of the inspiral is much longer than a single orbital period, and the entire process lasts for tens to hundreds of years.
As a result, although the instantaneous strength of the radiation is less significant when compared with binary merger events of supermassive objects, the accumulated signal can possibly be extracted using matched filtering. 
Also, the emitted GW is expected to locate in the sensitive band of space-based GW detector like LISA \cite{LISA:2017pwj}.

When calculating the energy loss of the orbital energy due to gravitational radiation absorbed by the primary BH or escaping to the spatial infinity, one needs to sum up the contributions from all possible radiation channels.
In general relativity (GR), astrophysical binary systems only emit gravitational waves with tensor polarizations, and the lowest radiative multipole moment is the quadrupole moment.
However, in alternative theories of gravity, additional emission channels may exist.
For example, in Brans-Dicke theory as well as some scalar-tensor theories, the additional scalar field activates the scalar dipole radiation~\cite{Alsing:2011er,Damour:1996ke,PhysRevD.56.785,PhysRevD.66.024040,Yagi:2015oca,Kuntz_2020}.
Moreover, the dynamics of the binary system will be modified by the presence of the dipole gravitational radiation \cite{1975ApJ196L59E, Berti:2004bd, Freire:2012mg, Barausse:2016eii}, which significantly deviates from that determined by GR.
In fact, even one assumes that the strength of the additional mode is insignificant compared with the tensorial ones, the extensive duration of the process is subjected to accumulated energy loss that eventually becomes observable.
The latter may demonstrate themselves in terms of imprint on the orbital dynamics, as well as the GW frequency \cite{Alsing:2011er, Berti:2012bp, Ramazanoglu:2016kul}.
Therefore, one may discriminate between different theories of gravity regarding the information on the additional degree of freedom extracted from empirical observations of EMRIs.
The present study is focused on the EMRI signals in the modified gravity, where an additional scalar field is non-minimally coupled to the metric.

For most EMRIs, the secondary object can be considered a test particle that moves in a fixed background generated by the central supermassive BH.
Therefore, the test particle follows a geodesic of the background on each time slice.
The geodesic parameters, namely, the particle's orbital energy and angular momentum, will change adiabatically due to the gravitational radiation.
In an alternative theory of gravity, the presence of an additional scalar degree of freedom, together with the metric fields governing the gravitational interaction, may violate the strong equivalence principle.
To be specific, an object's internal structure may depend on the local value of the scalar field.
Therefore, as the object changes its location, its internal structure varies, which, in turn, affects the object's motion.
Eardley~\cite{1975ApJ196L59E} first proposed an effective action to explore such an effect in Brans-Dicke's theory by assuming that the mass of a compact object is a general function of the scalar field.
In the so-called ``skeletonized approach''~\cite{1989ApJ346366W, Damour:1992we, Julie:2017ucp, Julie:2017rpw}, a compact object is treated as a point particle with the mass $m(\phi)$, and the action reads
\begin{equation}
\label{skeleton_action}
    S_{\rm p}=-\int m(\phi)ds=-\int m(\phi)\sqrt{g_{\mu\nu}\frac{dy_{\rm p}^\mu}{d\lambda}\frac{dy_{\rm p}^\nu}{d\lambda}}d\lambda\, ,
\end{equation}
where $s$ is the proper time of the particle along the worldline $y_{\rm p}^\mu(\lambda)$ at a given coordinate,
$\lambda$ is an affine parameter.
The effective action \eqref{skeleton_action} has been extensively employed to investigate the scalar dipole emission in alternative theories of gravity \cite{1989ApJ346366W, Damour:1996ke, Freire:2012mg, 2013Sci...340..448A, Kuntz_2020, Maselli:2020zgv, Niu:2019ywx,Jiang:2021htl}.
Furthermore, it has also been applied to the case of binaries consisting of hairy BHs \cite{Kanti:1995vq,Maeda:2009uy,Yunes:2011we,Sotiriou:2013qea,Sotiriou:2014pfa,Silva:2017uqg,Ventagli:2020rnx, Doneva:2017bvd, Guo:2020sdu, Yunes:2011aa, Sopuerta:2009iy,Pani:2011xj, Witek:2018dmd, Cardoso:2018zhm, Barack:2006pq, Dai:2021olt}.

In \cite{Maselli:2020zgv}, the authors investigated an EMRI in an alternative theory of gravity in which the central supermassive BH provides an approximate Schwarzschild background, and the secondary body with a scalar charge is described by the action \eqref{skeleton_action}.
It is shown that the scalar field emission has a significant cumulative effect on the dynamics of the EMRI.
In particular, the additional energy loss depends on the scalar charge of the secondary body and is relatively insensitive to the specific theory of gravity.
Ref. \cite{Maselli:2021men} extended the primary BH to a Kerr one but for chosen values of the mass ratio $10^{-5}$ and dimensionless spin $0.9$. 
The authors concluded that LISA could detect the scalar charge of the secondary body in EMRIs. 

Considering that astrophysical black holes have different scales and angular momenta, it is of great interest to further generalize the study of~\cite{Maselli:2021men} to the entire parameters space. 
We are interested in examining in which parameter range the accumulation of scalar field emission is more significant. 
In dynamical Chern-Simons gravity \cite{Jackiw:2003pm}, it was disclosed that the correction to the number of gravitational wave cycles depends on the mass ratio. 
With increasing mass ratio, the correction was found to initially rise to a maximum value and then fall back. 
Following this line of thought, it is meaningful to study the effects of various model parameters.
In particular, it would be intriguing to identify whether the impact on the relevant observables is monotonous or some optimal configuration can be encountered.
In this regard, we aim to explore the entire parameters space, consisting of the BH spin, mass ratio, and especially the charge of the scalar field. 
The relevant observable quantities include the EMRI radiation flux of different modes and the accumulated dephasing.
Among others, we expect to find a combined range of parameters, for which the accumulated effect of the scalar effect is possibly more substantial than that found in \cite{Maselli:2021men}.
It is also noted that instead of one year, we simulate the EMRI up to four years before the binary merge occurs.
This is because the difference in the energy emission exhibited in the EMRIs between GR and the modified gravity is mostly insignificant, and therefore, a more extensive observation period is desirable.
Besides, the LISA mission is proposed for a four-year duration, according to its science plan \cite{LISA:2017pwj,Barausse2020}.
Therefore, by running the simulations for its entire lifespan, one might give a reasonable estimation about the lightest scalar charge that is experimentally accessible by the LISA project.

This paper is organized as follows.
Sec. \ref{sec=BG} introduces the theoretical framework of the present study. 
Under appropriate approximations, in Sec. \ref{sec=teu} we present the equations of motion that govern the perturbations of the metric and scalar field.
The particular profiles of the latter are obtained via Green's function, which is constructed using the asymptotical solutions of the corresponding homogeneous differential equation. 
We present the numerical results in Sec. \ref{sec=result}, where we study the effects of various model parameters on the energy flux and dephasing.
The last section is devoted to further discussions and the conclusion.

\section{Setup of the background}\label{sec=BG}
The action is \cite{Maselli:2020zgv}
\begin{equation}\label{action}
S[g_{\mu\nu},\phi,\Psi]=S_0[g_{\mu\nu},\phi]+\alpha S_c[g_{\mu\nu},\phi]+S_m[g_{\mu\nu},\phi,\Psi],
\end{equation}
where
\begin{equation}
S_0=\int d^4x \frac{\sqrt{-g}}{16\pi}\left(R-\frac{1}{2}\partial_{\mu}\phi\partial^{\mu}\phi\right),
\end{equation}
$R$ is the Ricci scalar and $\phi$ is a scalar field.
The second term $\alpha S_c[g_{\mu\nu},\phi]$ denotes the non-minimal coupling between the metric $g_{\mu\nu}$ and scalar field $\phi$, and  $\alpha$ is a coupling constant with dimensions $[\alpha]=(mass)^n$ which characterizes the deviation from GR.
The action for the matter field $\Psi$ is $S_m[g_{\mu\nu},\phi,\Psi]$.
For EMRIs, the matter field is the secondary body with the mass $m(\phi)$ and its action is Eq. \eqref{skeleton_action}.

The equations of motion are obtained by varying the action \eqref{action}
\begin{eqnarray}\label{eom}
G_{\mu \nu}=R_{\mu \nu}-\frac{1}{2} g_{\mu \nu} R=T_{\mu \nu}^{\text{scal}}+\alpha T_{\mu \nu}^{c}+T_{\mu \nu}^{p},\label{eom1}\\
\square \phi+\frac{16 \pi \alpha}{\sqrt{-g}} \frac{\delta S_{c}}{\delta \phi}=16 \pi \int m^{\prime}(\phi) \frac{\delta^{(4)}\left(x-y_{p}(\lambda)\right)}{\sqrt{-g}} d \lambda,\label{eom2}
\end{eqnarray}
where $m^{\prime}(\phi)=dm(\phi)/d\phi$,
the stress-energy tensor of the scalar field is $T_{\mu \nu}^{\text{scal}}=\frac{1}{2} \partial_{\mu} \phi \partial_{\nu} \phi-\frac{1}{4} g_{\mu \nu}(\partial \phi)^{2}$,
$T_{\mu\nu}^c$ is the stress-energy of the coupling term and the stress-energy tensor of the test particle is given by
\begin{equation}
T_{\mathrm{p}}^{\alpha \beta}=8 \pi \int m(\phi) \frac{\delta^{(4)}\left(x-y_{p}(\lambda)\right)}{\sqrt{-g}} \frac{d y_{\rm p}^{\alpha}}{d \lambda} \frac{d y_{\rm p}^{\beta}}{d \lambda} d \lambda.
\end{equation}

Based on the discussions in Refs. \cite{Maselli:2020zgv,Maselli:2021men}, both the first two terms on the right-hand side of Eq. \eqref{eom1} and the second term on the left-hand side of Eq. \eqref{eom2} can be neglected.
Far away from all the sources, the scalar field approximates to $\phi=\phi_0+\frac{m_{\rm p} d}{r}+...$,
where $\phi_0$ stands for the background value of the scalar field,
$d$ is the dimensionless scalar charge of the body with mass $m(\phi_0)=m_{\rm p}$,
the ellipsis $...$ stands for higher-order infinitesimal quantity.
The $\phi$-dependent mass could be expanded as $m(\phi)=m_{\rm p}+m^{\prime}(\phi_0) \frac{m_{\rm p} d}{r}+...$.
Subsituting this expansion into Eq. \eqref{eom2},
one can obtain the relation $m^{\prime}(\phi_0)/m(\phi_0)=-d/4$.

Finally, to the lowest order, the perturbed field equations are
\begin{align}
G_{\mu\nu}&=T^{\rm p}_{\mu\nu}=8\pi m_{p}\int \frac{\delta^{(4)}(x-y_{p}(\lambda))}{\sqrt{-g}}
\frac{dy_{\rm p}^\alpha}{d\lambda}\frac{dy_{\rm p}^\beta}{d\lambda} d\lambda\label{eq:pertG},\\
  \square\phi&=-4\pi d\,m_{\rm p}\int \frac{\delta^{(4)}(x-y_{\rm p}(\lambda))}{\sqrt{-g}}d\lambda\,.\label{eq:pertphi}
\end{align}

Equation \eqref{eq:pertG} for the metric perturbation takes the same form as that in GR,
which is sourced by the inspiral of a secondary body with mass $m_{\rm p}$.
Thus the GW emission determined by the familiar tensor polarizations is the same as GR.
The additional equation \eqref{eq:pertphi} governs the scalar perturbation sourced by the scalar charge $d$.
Since the secondary body accelerates in a fixed background,
it generates scalar radiation.
Therefore, besides the familiar GW emission by tensor modes,
there exists scalar radiation in EMRIs.

\section{EMRIs in the Teukolsky framework}\label{sec=teu}

We focus on a quasi-circular orbital evolution of the secondary body spiraling into the supermassive Kerr BH adiabatically on the equatorial plane.
The Kerr BH provides the background on each time slice.
The test particle follows a geodesic in the background, and the geodesic parameters, i.e., the orbital energy and angular momentum of the particle,
change adiabatically due to gravitational radiation.
This section introduces the particle's orbital motion,
including the essential elements we will use in our following calculation,
derives and solves the perturbation equations in the Teukolsky framework and calculates the gravitational and scalar energy fluxes.
When numerically solving the perturbation equations,
we utilize the codes of BH Perturbation ToolKit \cite{BHPToolkit,Piovano:2020zin}.

We write the background Kerr BH spacetime in the Boyer-Lindquist coordinate,
\begin{equation} ds^2=-\left(1-\frac{2Mr}{\Sigma}\right)dt^2+\frac{\Sigma}{\Delta}dr^2-\frac{4Mar\sin^2\theta}{\Sigma}dtd\varphi+\Sigma d\theta^2+\frac{\sin^2\theta}{\Sigma}\left(\varpi^4-a^2\Delta\sin^2\theta\right)d\varphi^2,
\end{equation}
where $\Sigma\equiv r^2+a^2\cos^2\theta$, $\Delta\equiv r^2-2Mr+a^2$, $\varpi\equiv\sqrt{r^2+a^2}$,
$M$ is the mass of the central BH, the mass ratio is $q=m_{\rm p}/M$,
$a$ is the spin of the supermassive BH and satisfies $|a|\leq M$.
Without loss of generality, we assume $a$ is positive and aligned to the z-axis.
The inner and outer horizons of the BH are defined as $r_{\pm}=M^2\pm\sqrt{M^2-a^2}$. And here we define the tortoise coordinate $dr/dr^*=\Delta/(r^2+a^2)\equiv g(r)$.

\subsection{Orbital motion}
Geodesic motion in Kerr spacetime is completely integrable; this leads to three constants of motion: the specific energy $E$, the angular momentum $L$, and the Carter constant $Q$.
The geodesic equations are
\begin{eqnarray}
m_{\rm p}\Sigma_{w} \frac{d t_{w}}{d \tau} &=&E \frac{\varpi^{4}}{\Delta}+a L\left(1-\frac{\varpi^{2}}{\Delta}\right)-a^{2} E\sin^{2}\theta, \label{timequa}\\
m_{\rm p}\Sigma_{w} \frac{d r_{w}}{d \tau} &=&\pm \sqrt{V_{r}\left(r_{\rm p}\right)}, \\
m_{\rm p}\Sigma_{w} \frac{d \theta_{w}}{d \tau} &=&\pm \sqrt{V_{\theta}\left(\theta_{\rm p}\right)}, \\
m_{\rm p}\Sigma_{w} \frac{d \varphi_{w}}{d \tau} &=&a E\left(\frac{\varpi^{2}}{\Delta}-1\right)-\frac{a^{2} L}{\Delta}+ L \csc ^{2} \theta,\label{anglequa}
\end{eqnarray}
where the subscript $w$ stands for a function evaluated on the worldline of the secondary body, the radial and polar potentials are
\begin{eqnarray}
&V_{r}(r)& =\left(E \varpi^{2}-a L\right)^{2}-\Delta\left(r^{2}+\left(L-a E\right)^{2}+Q\right), \\
&V_{\theta}(\theta)& = Q-L^{2} \cot ^{2} \theta-a^{2}\left(1-E^{2}\right) \cos ^{2} \theta.
\end{eqnarray}

In the adiabatic approximation, for a quasi-circular orbit on the equatorial plane,
the coordinates $r$ and $\theta$ are considered as constants,
then Eqs. \eqref{timequa}  and \eqref{anglequa} are the remaining equations.
For this orbit, the conserved constants $E$ and $L$ are expressed as \cite{Detweiler687D}
\begin{eqnarray}\label{orbitE}
	E/m_{\rm p}&=& \frac{r_{\rm p}^{3 / 2}-2 M r_{\rm p}^{1 / 2} \pm a M^{1 / 2}}{r_{\rm p}^{3 / 4}\left(r_{\rm p}^{3 / 2}-3 M r_{\rm p}^{1 / 2} \pm 2 a M^{1 / 2}\right)^{1 / 2}},    \\
	L/m_{\rm p}&=& \frac{\pm M^{1/2}(r_{\rm p}^{2}\mp 2aM^{1/2} r_{\rm p}^{1 / 2} + a^2)}{r_{\rm p}^{3 / 4}\left(r_{\rm p}^{3 / 2}-3 M r_{\rm p}^{1 / 2} \pm 2 a M^{1 / 2}\right)^{1 / 2}}.
\end{eqnarray}
The orbital angular frequency, measured by a distant observer, is given by \cite{Detweiler687D}
\begin{equation}\label{orbitF}
\Omega \equiv \frac{d\varphi}{dt}=\frac{\pm M^{1 / 2}}{r_{\rm p}^{3 / 2} \pm a M^{1 / 2}}.
\end{equation}
In Eqs. \eqref{orbitE} and \eqref{orbitF} the upper sign refers to the case co-rotating with the central BH while the lower one refers to the counter-rotating case.
In the following discussion, we consider the co-rotating motion without loss of generality.

\subsection{Metric perturbations}

We use the Teukolsky formalism to calculate the energy flux of GW emission.
At spatial infinity, the metric perturbation is related to the Newman-Penrose quantity $\psi_4$ as
\begin{equation}
\psi_4=\frac{1}{2}\frac{\partial^2}{\partial t^2}(h_+-i h_{\times}).
\end{equation}
The Newman-Penrose variable $\psi_4$ can be expanded as
\begin{equation}
\psi_{4}=\rho^{4} \sum_{\ell=2}^{\infty} \sum_{m=-\ell}^{\ell} \int_{-\infty}^{\infty} d \hat{\omega} R_{\ell m \hat{\omega}}(r)_{-2} S_{\ell m}^{a \hat{\omega}}(\theta) e^{i(m \varphi-\hat{\omega} \hat{t})},
\end{equation}
where $\rho=(r-i a\cos\theta)^{-1}$, $_{-2} S_{\ell m}^{a \hat{\omega}}(\theta)$ is the $s=-2$ spin weighted orthonormal spheroidal harmonics, and its eigenvalue is $\lambda_G$.
The radial function $R_{\ell m \hat{\omega}}$ obeys the radial Teukolsky equation
\begin{equation}\label{radialGR}
\Delta^{2} \frac{d}{d r}\left(\frac{1}{\Delta} \frac{d R_{\ell m \omega}}{d r}\right)-V_G(r) R_{\ell m \hat{\omega}}(r)=\mathcal{T}^T_{\ell m \hat{\omega}},
\end{equation}
where the source term $\mathcal{T}^T_{\ell m \hat{\omega}}$ is constructed from the stress-energy tensor in Eq. \eqref{eq:pertG}~\cite{Hughes:1999bq,Chrzanowski:1976jy}.
The potential $V_G(r)$ takes the form
\begin{eqnarray}
	V_G(r) &=&-\frac{K^{2}+4 i(r-1) K}{\Delta}+8 i \hat{\omega} r+\lambda_G,
\end{eqnarray}
where $K=\left(r^{2}+a^{2}\right) \hat{\omega}-a m$.
The gravitational energy flux at the horizon and at infinity are given by \cite{Teukolsky:1973ha,Hughes:1999bq}
\begin{eqnarray}
	\dot{E}_{T}^{H}&=&\sum_{\ell=2}^{\infty} \sum_{m=1}^{\ell} \alpha_{\ell m} \frac{\left|\mathcal{A}_{\ell m \hat{\omega}}^{\infty}\right|^{2}}{2\pi(m \Omega)^{2}},\\
	\dot{E}_{T}^{\infty}&=&\sum_{\ell=2}^{\infty} \sum_{m=1}^{\ell} \frac{\left|\mathcal{A}_{\ell m \hat{\omega}}^{H}\right|^{2}}{2\pi(m \Omega)^{2}},
\end{eqnarray}
where the subscript ``$T$" stands for tensor modes, superscripts ``$H$" and ``$\infty$" are for horizon and infinity, respectively.
The coefficients are
\begin{equation}
\alpha_{\ell m}=\frac{256\left(2 r_{+}\right)^{5} \hat{\kappa}\left(\hat{\kappa}^{2}+4 \epsilon^{2}\right)\left(\hat{\kappa}^{2}+16 \epsilon^{2}\right)(m \Omega)^{3}}{\left|C_{\ell m}\right|^{2}},
\end{equation}
\begin{equation}
\begin{aligned}
\left|C_{\ell m}\right|^{2} &=\left[\left(\lambda_G+2\right)^{2}+4 \hat{a}(m \Omega)-4 a^{2}(m \Omega)^{2}\right]\times\left[\lambda_G^{2}+36 m a(m\Omega)-36a^{2}(m \Omega)^{2}\right] \\
&+\left(2 \lambda_G+3\right)\left[96 a^{2}(m \Omega)^{2}-48 m a(m \Omega)\right]+144(m \Omega)^{2}\left(1-a^{2}\right)
\end{aligned}
\end{equation}
with $\epsilon=\sqrt{1-a^2}/(4r_+)$ and $\hat{\kappa}=\hat{\omega}-ma/(2r_+)$.
The function $\mathcal{A}_{\ell m \hat{\omega}}^{H,\infty}$ comes from the solution of Eq. \eqref{radialGR},
\begin{eqnarray}
&Z_{\ell m \hat{\omega}}^{H, \infty}=\delta(\hat{\omega}-m \Omega) \mathcal{A}_{\ell m \hat{\omega}}^{H, \infty},\\
&R_{\ell m \hat{\omega}}\left(r\rightarrow r_{+}\right)=Z_{\ell m \hat{\omega}}^{\infty} \Delta^{2} e^{-i \hat{\kappa} r^{*}}, \\
&R_{\ell m \hat{\omega}}(r\rightarrow\infty)=Z_{\ell m \hat{\omega}}^{H} r^{3} e^{i \hat{\omega} r^{*}}.
\end{eqnarray}

\subsection{Scalar perturbations}
To solve Eq. \eqref{eq:pertphi},
we expand the scalar perturbation $\phi$ in scalar spheroidal harmonics as
\begin{equation}
\phi(t, r, \theta, \varphi)=\sum_{\ell, m} \int d\hat{\omega}~e^{i(m \varphi-\hat{\omega}t)}\frac{X_{\ell m\hat{\omega}}(r)}{\sqrt{r^{2}+a^{2}}} {}_0 S_{\ell m}(\theta),
\end{equation}
where ${}_0 S_{\ell m}(\theta)$ is the $s=0$ orthnormal spheroidal harmonics with the eigenvalue $\lambda_s$.

The radial perturbation of the scalar field reads
\begin{equation}\label{radialsca}
\left[\frac{d^{2}}{d r_{*}^{2}}+V_{s}(r)\right] X_{\ell m \hat{\omega}}(r)=\frac{\Delta}{\left(r^{2}+a^{2}\right)^{3 / 2}} \mathcal{T}^s_{\ell m \hat{\omega}},
\end{equation}
and the effective potential
\begin{equation}
V_{s}=\left(\hat{\omega}-\frac{a m}{\sigma}\right)^{2}-\frac{\Delta}{\sigma^{4}}\left[\lambda_s ~\sigma^{2}+2 M r^{3}+a^{2}\left(r^{2}-4 M r+a^{2}\right)\right],
\end{equation}
where $\sigma=r^2+a^2$, and $\mathcal{T}^s_{\ell m\hat{\omega}}$ is constructed from the source term on the right hand side of Eq. \eqref{eq:pertphi}.

Following the discussion of \cite{Piovano:2020zin},
we rewrite the homogeneous radial perturbation equation
\begin{equation}\label{SNequa}
\frac{d^{2}}{~dr^{2}}X_{\ell m \hat{\omega}}+F(r)\frac{d}{dr}X_{\ell m \hat{\omega}}+G(r) X_{\ell m \hat{\omega}}=0
\end{equation}
with $F(r)\equiv g'(r)/g(r)$ and $G(r)\equiv V_s/g(r)^2$.
Eq. \eqref{SNequa} has two linearly independent solutions which satisfy pure ingoing boundary condition at the horizon and pure outgoing boundary condition at infinity,
\begin{equation}
\begin{array}{ll}
X_{\ell m \hat{\omega}}^{\rm in} \sim \begin{cases}e^{-i \hat{\kappa} r^{*}}, & r \rightarrow r_{+}, \\
A_{\ell m \hat{\omega}}^{\rm out} e^{i \hat{\omega} r^{*}}+A_{\ell m \hat{\omega}}^{\rm in} e^{-i \hat{\omega} \hat{r}^{*}}, & r \rightarrow \infty,
\end{cases} \\
X_{\ell m \hat{\omega}}^{\rm up} \sim \begin{cases}C_{\ell m \hat{\omega}}^{\rm out} e^{i \hat{\kappa}r^{*}}+C_{\ell m \hat{\omega}}^{\rm in} e^{-i \hat{\kappa} r^{*}}, & r \rightarrow r_{+}, \\
e^{i \hat{\omega} r^{*}}, & r \rightarrow \infty,
\end{cases}
\end{array}
\end{equation}
where $\hat{\kappa}(\hat{\omega})\equiv \hat{\omega}-ma/(2r_+)$.
Through the Green's function method, the solution is expressed as
\begin{equation}
X_{\ell m \hat{\omega}}(r) =X_{\ell m \hat{\omega}}^{\rm up}(r) \int_{r_{+}}^{r} ~d r^{\prime} \frac{X_{\ell m \hat{\omega}}^{\rm in}\left(r^{\prime}\right) \mathcal{T}^s_{\ell m \hat{\omega}}\left(r^{\prime}\right)}{W_{r}}+X_{\ell m \hat{\omega}}^{\rm in}(r) \int_{r}^{\infty} d r^{\prime} \frac{X_{\ell m \hat{\omega}}^{\rm up}\left(r^{\prime}\right) \mathcal{T}^s_{\ell m \hat{\omega}}\left(r^{\prime}\right)}{W_{r}},
\end{equation}
where the Wronskian is $W_r\equiv X^{\rm in}_{\ell m\hat{\omega}}dX^{\rm up}_{\ell m\hat{\omega}}/dr^*-X^{\rm up}_{\ell m\hat{\omega}}dX^{\rm in}_{\ell m\hat{\omega}}/dr^*$.
Since the inhomogeneous solution is purely ingoing at the horizon and outgoing at infinity,
it can be rewritten as
\begin{eqnarray}
	X_{\ell m \hat{\omega}}\left(r \rightarrow r_{+}\right) &=\mathcal{Z}_{\ell m \hat{\omega}}^{\infty} e^{-i \hat{\kappa} r^{*}}, \\
    X_{\ell m \hat{\omega}}(r \rightarrow \infty) &=\mathcal{Z}_{\ell m \hat{\omega}}^{H} e^{i \hat{\omega} r^{*}},
\end{eqnarray}
where the coefficients are
\begin{equation}
\mathcal{Z}_{\ell m \hat{\omega}}^{H, \infty}=-4\pi dq\frac{X_{\ell m \hat{\omega}}^{\rm in,up}\left(r_{\rm p}\right)}{W_r\ u^{t}}\ \frac{{}_0 S_{\ell m}^{*}(\pi / 2)}{\sqrt{r_{\rm p}^{2}+a^{2}}},
\end{equation}
${}_0S_{\ell m}(\theta)^*$ is the complex conjugation of ${}_0S_{\ell m}(\theta)$, $u^\mu$ is the four velocity of the particle, and $r_{\rm p}$ is the orbital radius.
The scalar energy flux at the horizon and at infinity is \cite{Teukolsky:1973ha,Yunes:2011aa}
\begin{eqnarray}
\dot{E}_{s}^{H}&=&\frac{1}{16\pi}\sum_{\ell=1}^{\infty} \sum_{m=-\ell}^{\ell} m ~\Omega ~\hat{\kappa}(m\Omega)\left|\mathcal{Z}_{l m \omega}^{\infty}\right|^{2},\\
\dot{E}_{s}^{\infty}&=&\frac{1}{16\pi}\sum_{\ell=1}^{\infty} \sum_{m=-\ell}^{\ell} m^2 ~\Omega^2\left|\mathcal{Z}_{\ell m \omega}^{H}\right|^{2}.
\end{eqnarray}

\section{Numerical results}\label{sec=result}

This section presents the numerical results.
In our simulation, we sum all the multipole contributions up to $\ell=18$.
The total energy flux is given by
\begin{equation} \mathcal{F}_{\rm tot}=\dot{E}_{GR}+\delta\dot{E}_{d}=\dot{E}^H_G+\dot{E}^\infty_G+\dot{E}^H_s+\dot{E}^\infty_s,
\end{equation}
where $\dot{E}_{GR}$ is for the total energy flux of tensor modes and $\delta\dot{E}_d$ is for the energy flux of scalar emission.

\begin{figure}[thbp]
\center{
\includegraphics[scale=0.54]{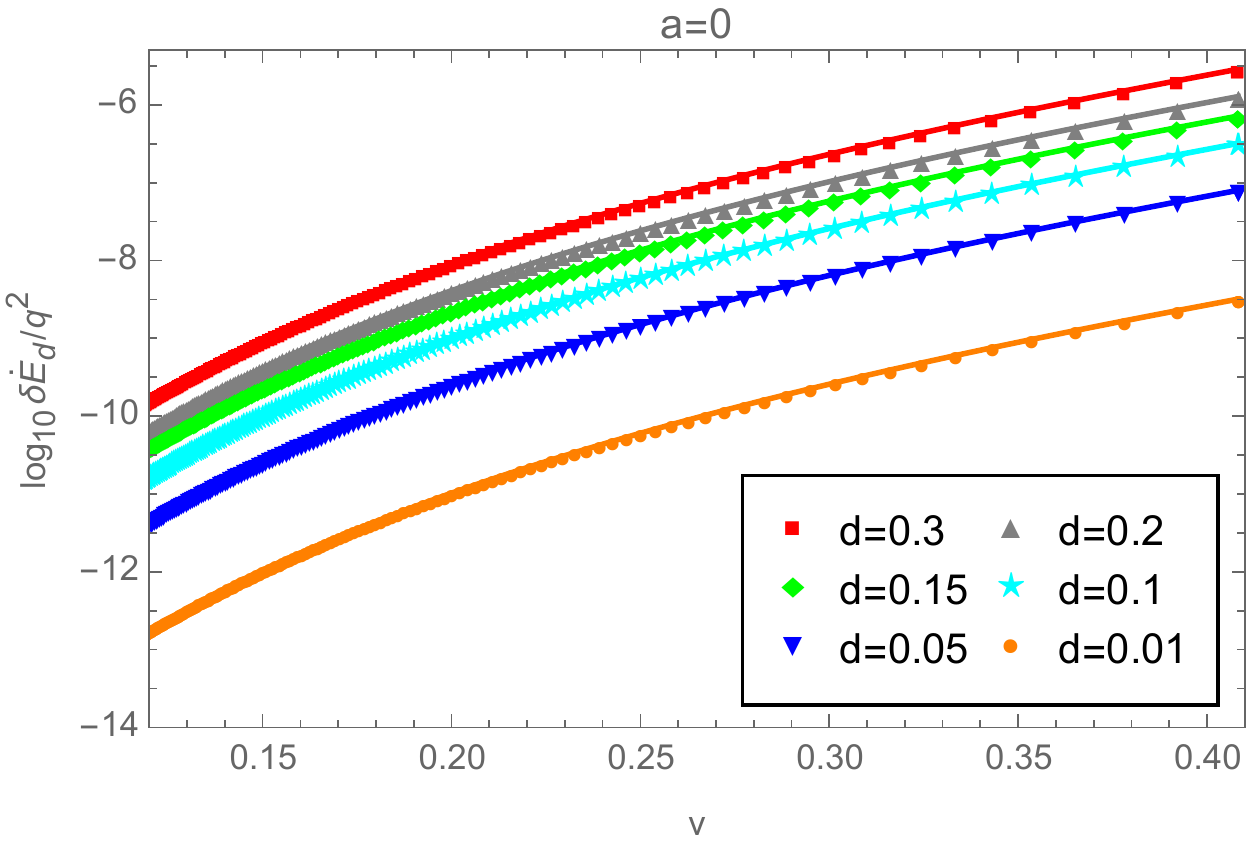}
\includegraphics[scale=0.54]{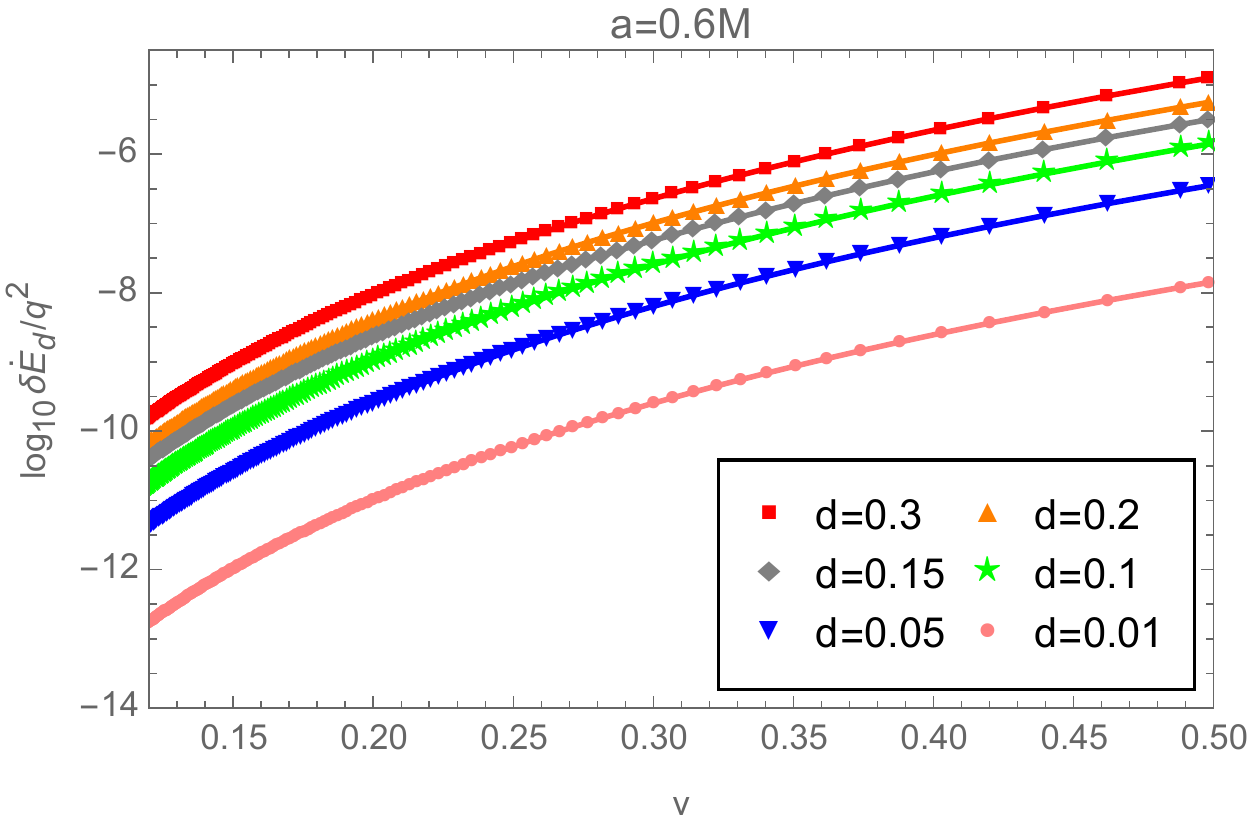}
\includegraphics[scale=0.54]{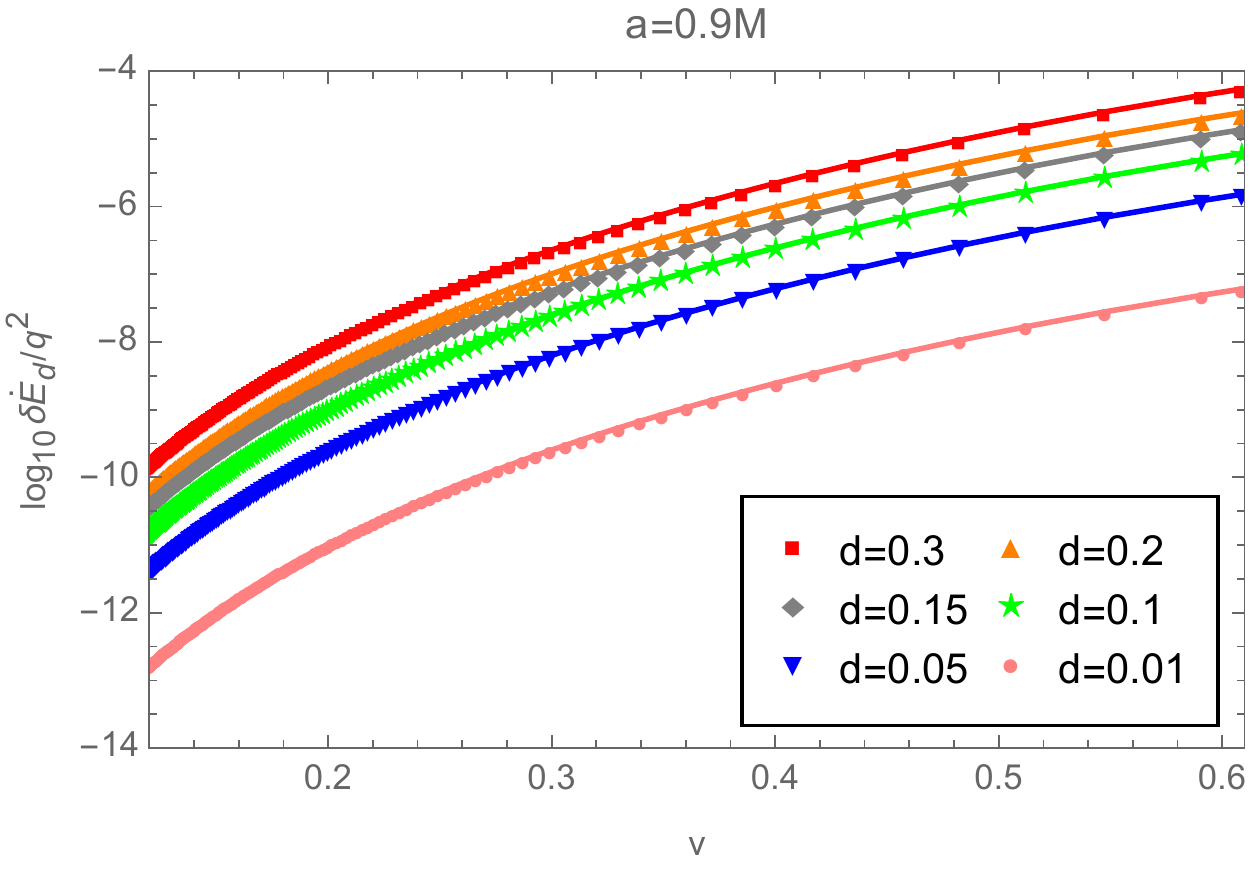}
\includegraphics[scale=0.54]{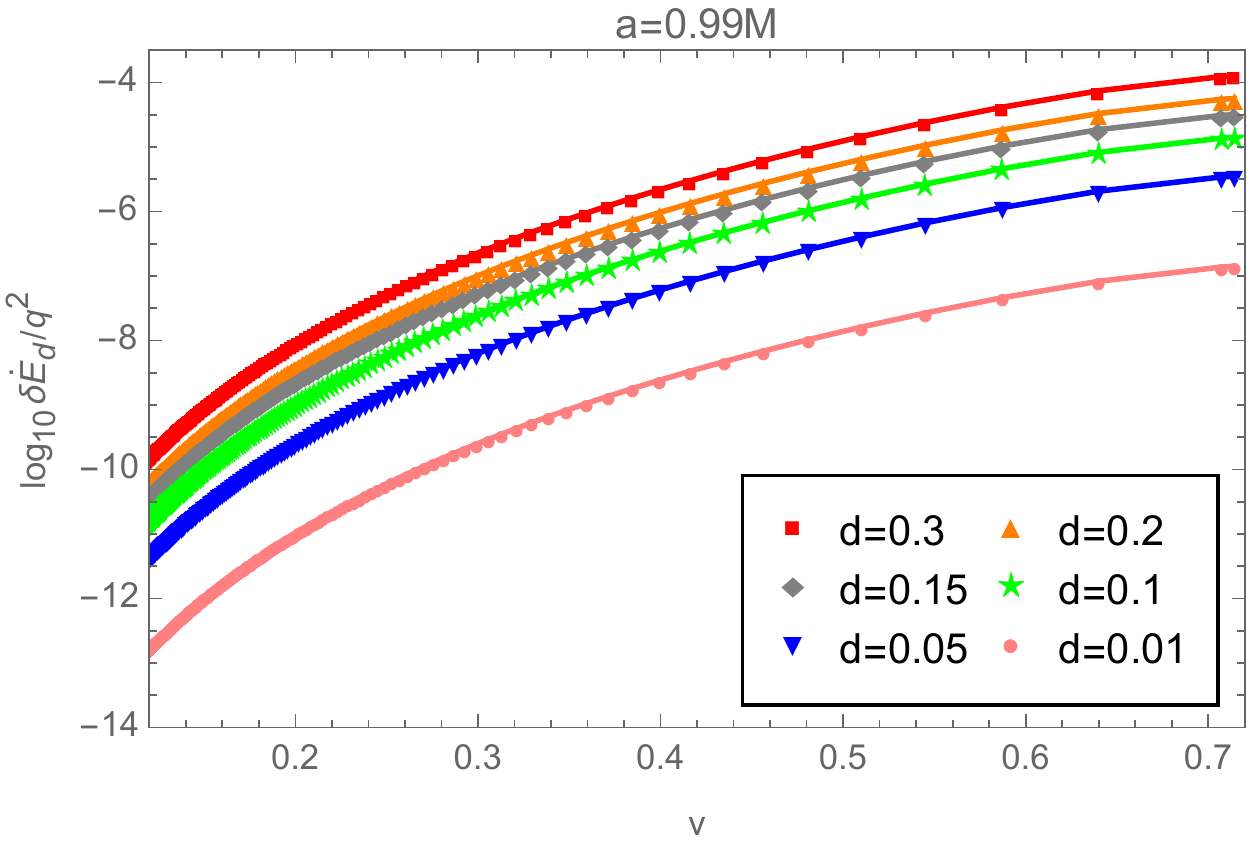}
\caption{The normalized value of the scalar energy flux $\delta\dot{E}_d$ as the function of the radial orbit velocity $v=(M \Omega)^{1/3}$ with different scalar charge $d$.
We take $a=0,\ 0.6M,\ 0.9M$ and $0.99M$. }\label{scalarflux}}
\end{figure}

Fig. \ref{scalarflux} shows the dependence of the normalized scalar energy flux $\delta\hat{E}_d/q^2$ on the orbital velocity $v=(M \Omega)^{1/3}$, the mass ratio $q$ is set to be $2\times 10^{-5}$, the region of the horizontal $v$-axis corresponds to the orbital radius starting from $r=80M$ and ending at the innermost stable circular orbit (ISCO).
We see that the scalar energy flux increases monotonously as the secondary object inspiralling into the central BH until it plunges into the ISCO.
For a fixed velocity of the secondary body,
as the scalar charge $d$ increases, the scalar energy flux increases.
Note that for an EMRI with a high spinning primary BH,
as the secondary body inspirals near the ISCO,
the scalar energy flux increases significantly.
For example, in the case of the primary BH with the spin $a=0.99M$, the strength of the scalar energy flux is almost two orders of magnitude larger than that with $a=0$.

\begin{figure}[thbp]
\center{
\includegraphics[scale=0.54]{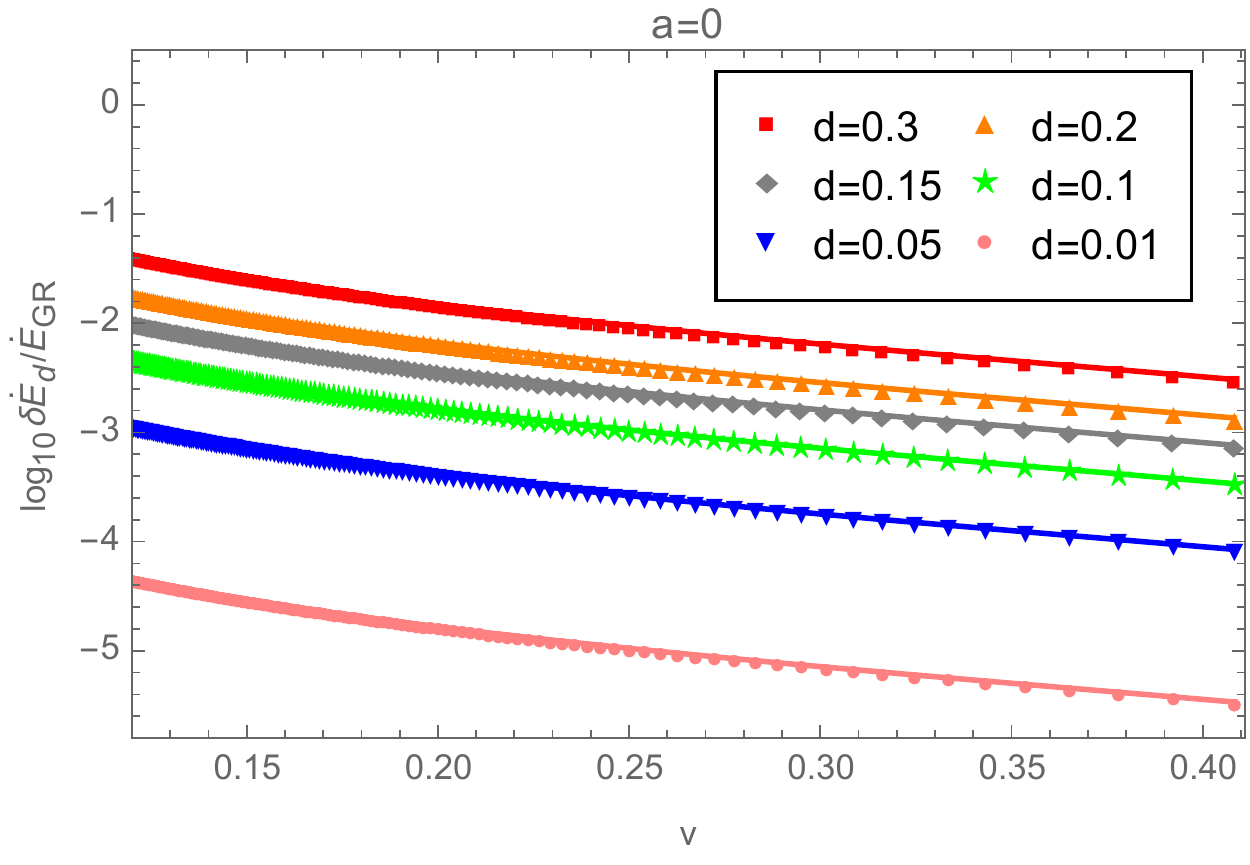}
\includegraphics[scale=0.54]{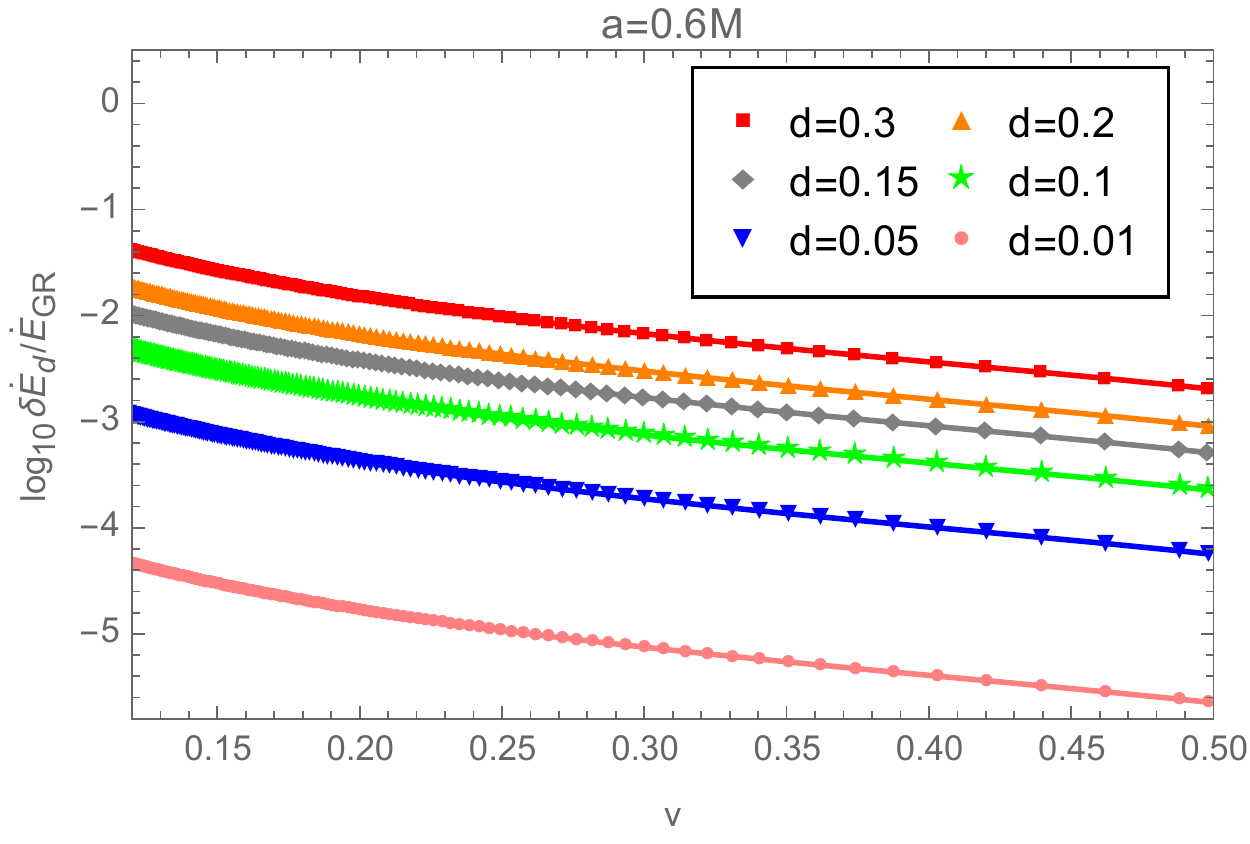}
\includegraphics[scale=0.54]{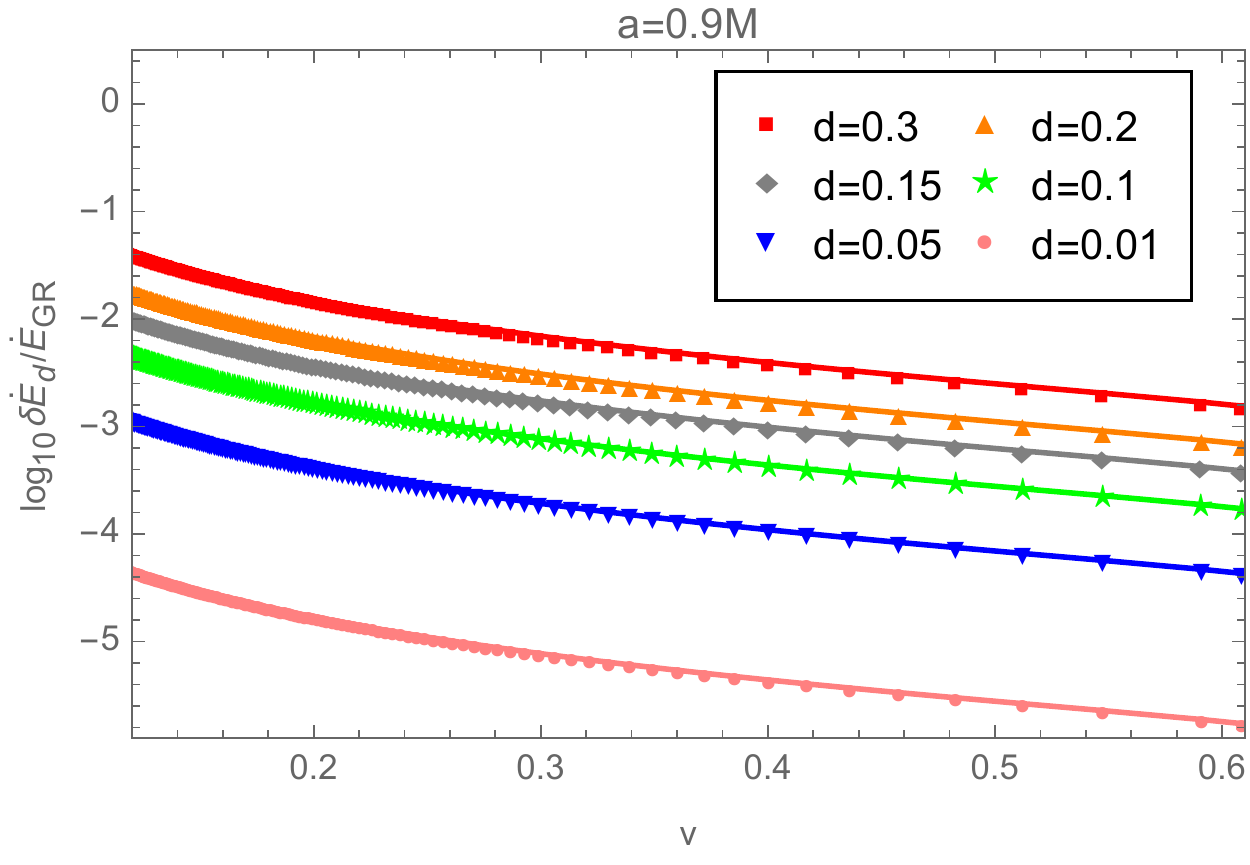}
\includegraphics[scale=0.54]{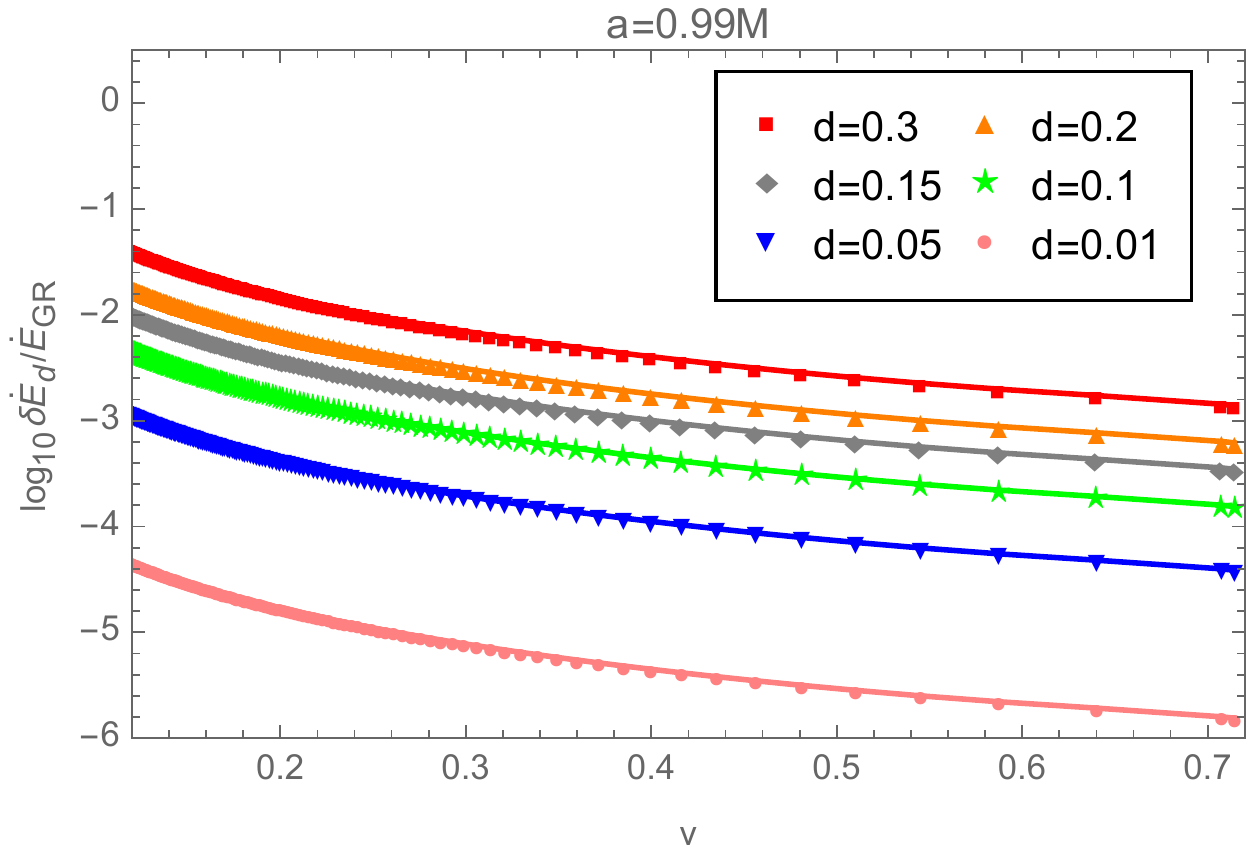}
\caption{The relative ratio $\delta\dot{E}_d/\dot{E}_{GR}$ as the function of the radial orbital velocity with different scalar charge $d$. We take $a=0,\ 0.6M,\ 0.9M$ and $0.99M$.}\label{GraScalux}}
\end{figure}

In Fig. \ref{GraScalux}, we show the relative ratio $\delta\dot{E}_d/\dot{E}_{GR}$ as a function of the orbital velocity $v=(M \Omega)^{1/3}$.
A more significant scalar charge increases the relative ratio because the source term of the scalar perturbation in Eq. \eqref{eom2} is proportional to the scalar charge $d$.
For different spin values $a$ of supermassive BHs,
the ratio $\delta\dot{E}_d/\dot{E}_{GR}$ decreases as the orbital velocity $v=(M \Omega)^{1/3}$, this is because gravitational energy flux grows more rapidly than the scalar energy flux in the high-frequency region.
For small scalar charges, the relative ratio is small; for example, when the scalar charge $d=0.01$, the relative ratio is smaller than $10^{-4}$.
However, the secondary body can orbit the central BH for more than $10^5$ cycles before the coalescence,
so the tiny deviation accumulates and eventually becomes detectable by the future space-based GW detector.

To show this point, we study the dephasing of the gravitational waveform caused by the additional energy loss during inspirals.
The obital angular frequency $\Omega$ is
 \begin{equation}\label{orbittime1}
\frac{d\varphi}{dt}=\Omega(r(t)),
 \end{equation}
where $\varphi$ is referred to as the orbital phase $\varphi_{\rm orb}$
and the GW frequency $f=\Omega/\pi$.
We take four years before the merger for the observation time,
\begin{equation}\label{sol-fre}
	T_{\text{obs}}=\int^{f_{\rm max}}_{f_{\rm min}}\frac{1}{\dot{f}}df=4\ \text{years},
\end{equation}
with
\begin{equation}\label{fmaxmin}
    f_{\text{max}}=min(f_{\rm ISCO},f_{\rm up}),~~~~~~f_{\text{min}}=max(f_{\rm low},f_{\rm start}),
\end{equation}
where $f_{\rm ISCO}$ is the frequency that the test particle reaches the ISCO at $t=t_{\rm end}$ and $f_{\rm start}$ is the initial frequency at $t=0$,
$f_{\rm low}=10^{-4}$Hz and $f_{up}=1$ Hz. 
The time derivative of the frequency is
\begin{equation}
\dot{f}=\frac{df}{dr}\frac{dr}{dt},
\end{equation}
where $dr/dt$ is determined from the energy conservation
 \begin{equation}\label{orbittime}
 \frac{d r}{dt}=-\mathcal{F}_{\rm tot}(t)\left(\frac{d E}{dr}\right)^{-1}.
 \end{equation}
Since
\begin{equation}
\frac{d \varphi}{d f}=\frac{\pi f}{\dot{f}},
\end{equation}
so the total phase before the merger can also be calculated by
\begin{equation}\label{phase-end}
\varphi_{\rm orb}(t_{\rm end})=\pi\int_{f_{\rm start}}^{f_{\rm max }} \frac{f}{\dot{f}} d f.
\end{equation}

If the scalar wave emission vanishes, 
$\mathcal{F}_{\rm tot}$ is determined entirely by the gravitational radiation of tensor modes, 
then the model reduces to its counterpart in GR and Eq. \eqref{phase-end} gives the phase $\varphi_{\rm orb}^{GR}(t_{\rm end})$ in GR.
To discuss the effect of scalar field on dephasing, 
we separate the total orbital phase into two parts $\varphi_{\rm orb}(t)=\varphi_{\rm orb}^{GR}(t)-\delta\varphi_{\rm orb}(t)$.
The quantity $\delta\varphi_{\rm orb}(t)$ denotes the deviation in the orbital phase from GR.
Compared with the case in GR,
the additional scalar emission increases the energy flux and causes the orbit to shrink faster,
so the small object experiences fewer cycles before the coalescence.
For the dominant mode, the dephasing accumulated till the coalescence due to the scalar wave emission is $\delta\varphi_{GW}(t_{\rm end})=2(\varphi_{\rm orb}^{GR}(t_{\rm end})-\varphi_{\rm orb}(t_{\rm end}))$.

\begin{figure}[thbp]
\center{
\includegraphics[scale=0.54]{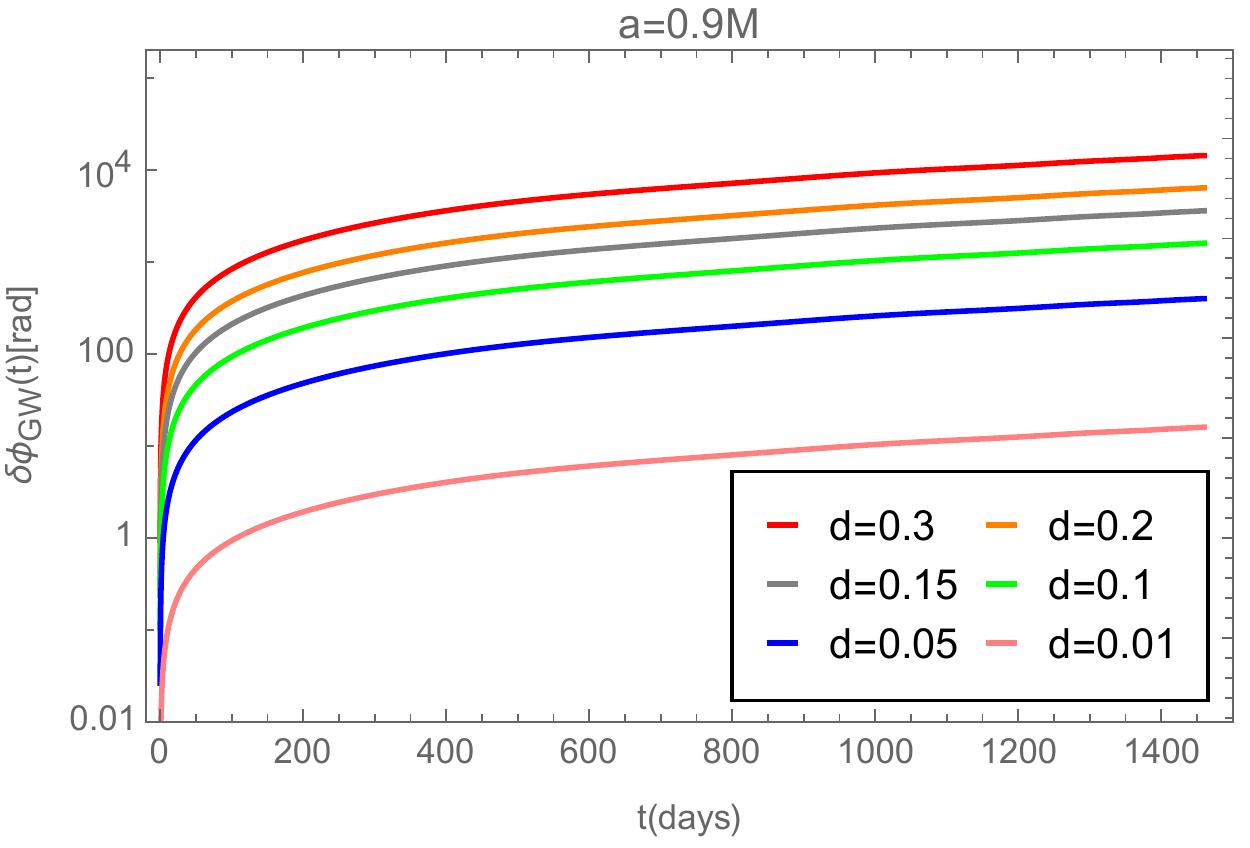}
\includegraphics[scale=0.54]{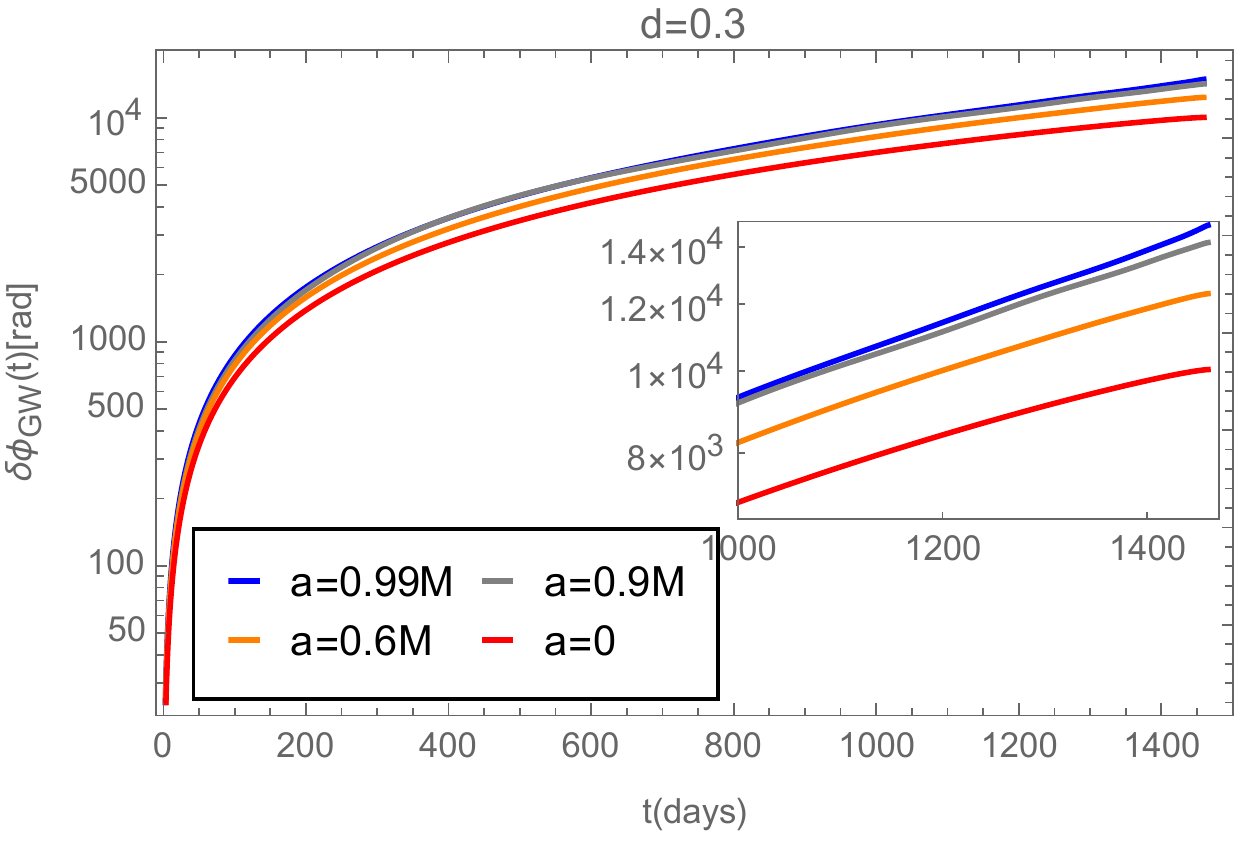}
\caption{Four years observation before the plunge.
Left: fixing $a=0.9M$, the dephasing $\delta\varphi_{GW}$ as the function of the time with different scalar charge $d$; Right: by setting $d=0.3$, the dephasing $\delta\varphi_{GW}$ as the function of the time with different BH spin $a$. Both of the data are under the assumption that $m_{\rm p}=10M_{\odot}$ and $q=2\times 10^{-5}$.
}\label{phaseda}}
\end{figure}

Fig. \ref{phaseda} shows the dephasing $\delta\varphi_{GW}(t)$, measured in radians, for different scalar charge $d$ and spin $a$ of the supermassive BH.
In the left figure, we take $a=0.9M$ and show that the dephasing grows with time.
For all the values of $d$ in the left figure, the final dephasing $\delta\varphi_{GW}(t_{\rm end})$ is bigger than 1 radian which are detectable.
In the right panel of Fig. \ref{phaseda}, 
by setting the scalar charge $d=0.3$ and varying the spin of the central BH, 
we show the dephasing $\delta\varphi_{GW}(t)$ as a function of time.
The dephasing is cumulative with time for EMRIs, and it reaches the maximum value at $r=r_{\rm ISCO}$.
A central BH with a bigger spin has a smaller ISCO,
and the test particle can spiral closer to the central BH where the radiated energy flux is larger (see Fig. \ref{GraScalux}).
Therefore, for the same observation time before the plunge, 
the final value of $\delta\varphi_{GW}(t_{\rm end})$ is larger for bigger $a$ as shown in Fig. \ref{phaseda}.
This means that the observation of EMRIs with a larger-spin Kerr BH can detect the scalar field easier.

\begin{figure}[thbp]
\center{
\includegraphics[scale=0.54]{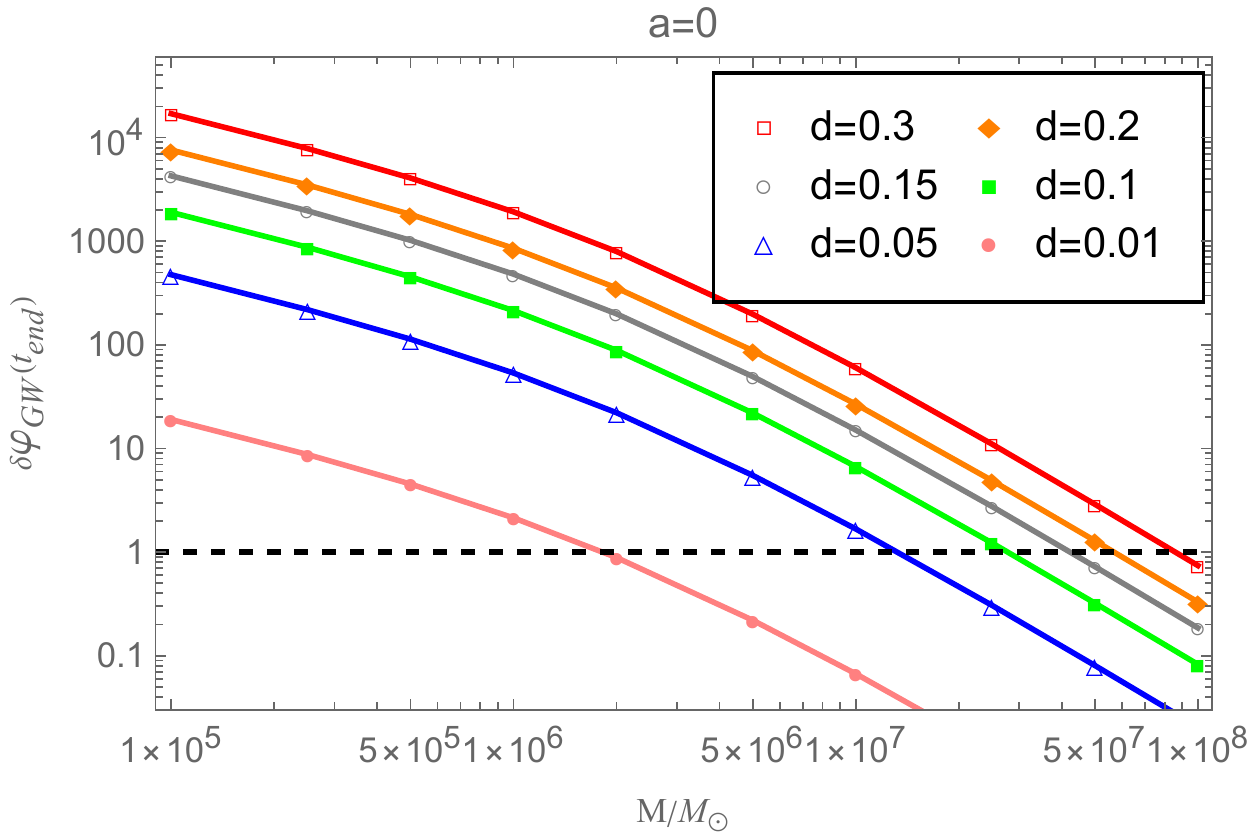}
\includegraphics[scale=0.54]{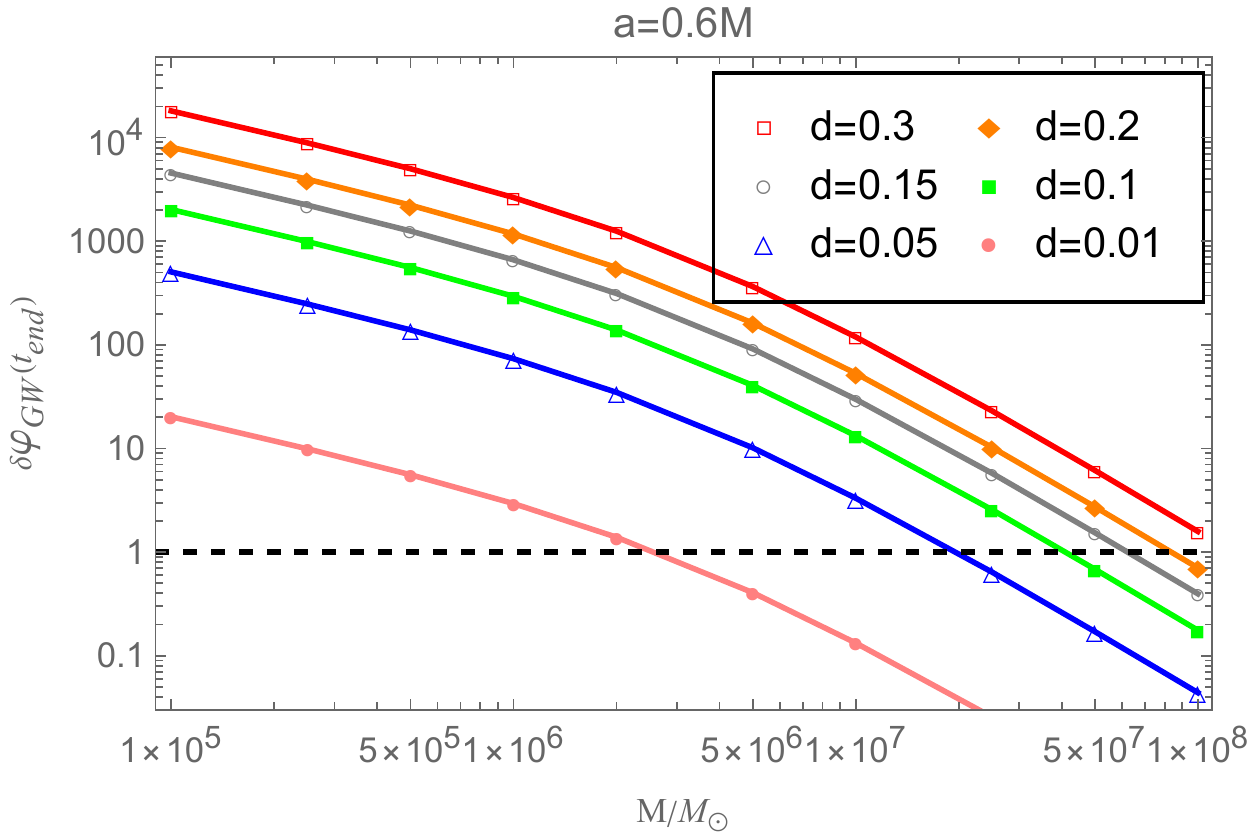}
\includegraphics[scale=0.54]{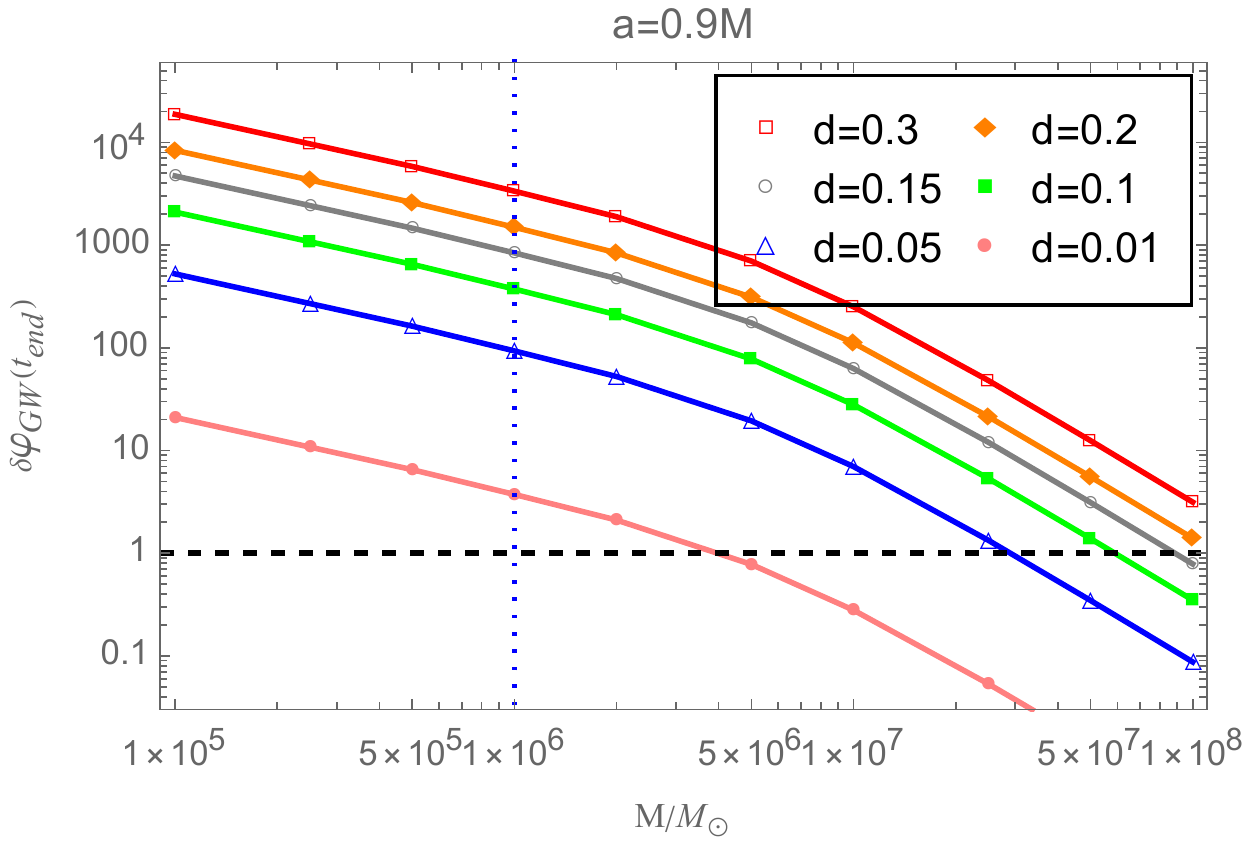}
\includegraphics[scale=0.54]{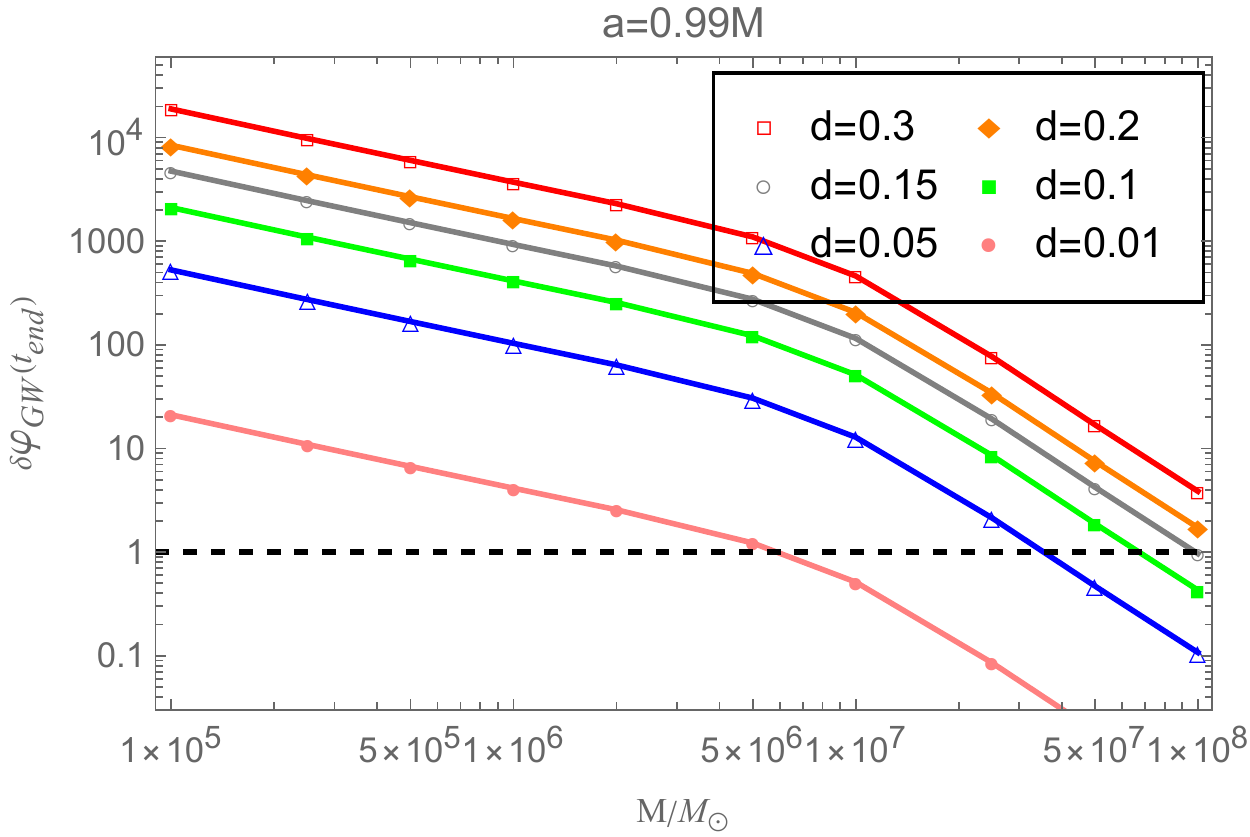}
\caption{The dephasing $\delta\varphi_{GW}(t_{\rm end})$ with $m_{\rm p}=10M_{\odot},\ M\in [10^5,10^8]M_{\odot}$ under different value of the scalar charge $d$, for example, we set $a=0,\ 0.6M,\ 0.9M,\ 0.99M$ correspondingly.
The black dashed line represents $\delta\varphi_{GW}(t_{\rm end})=1$ radian.}\label{cycles}}
\end{figure}

\begin{figure}[thbp]
\center{
\includegraphics[scale=0.42]{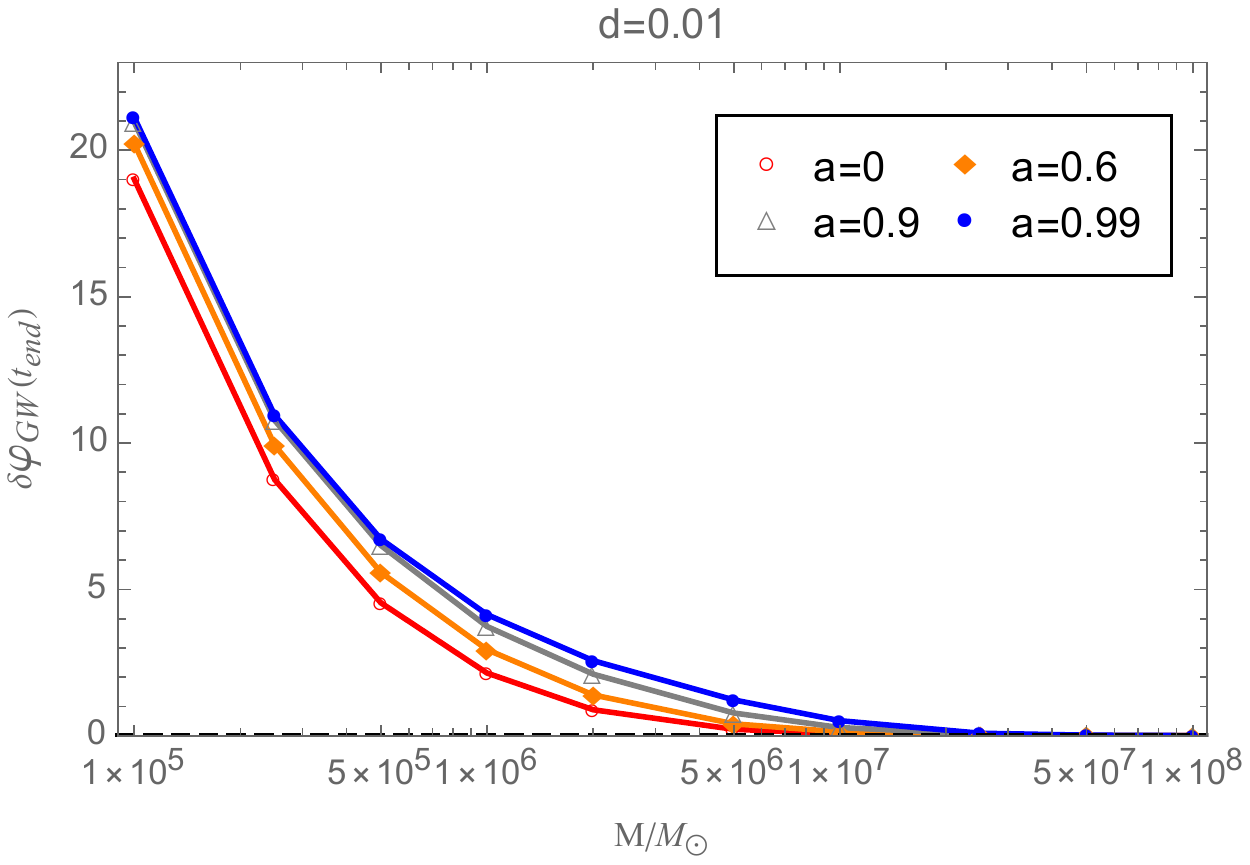}
\includegraphics[scale=0.42]{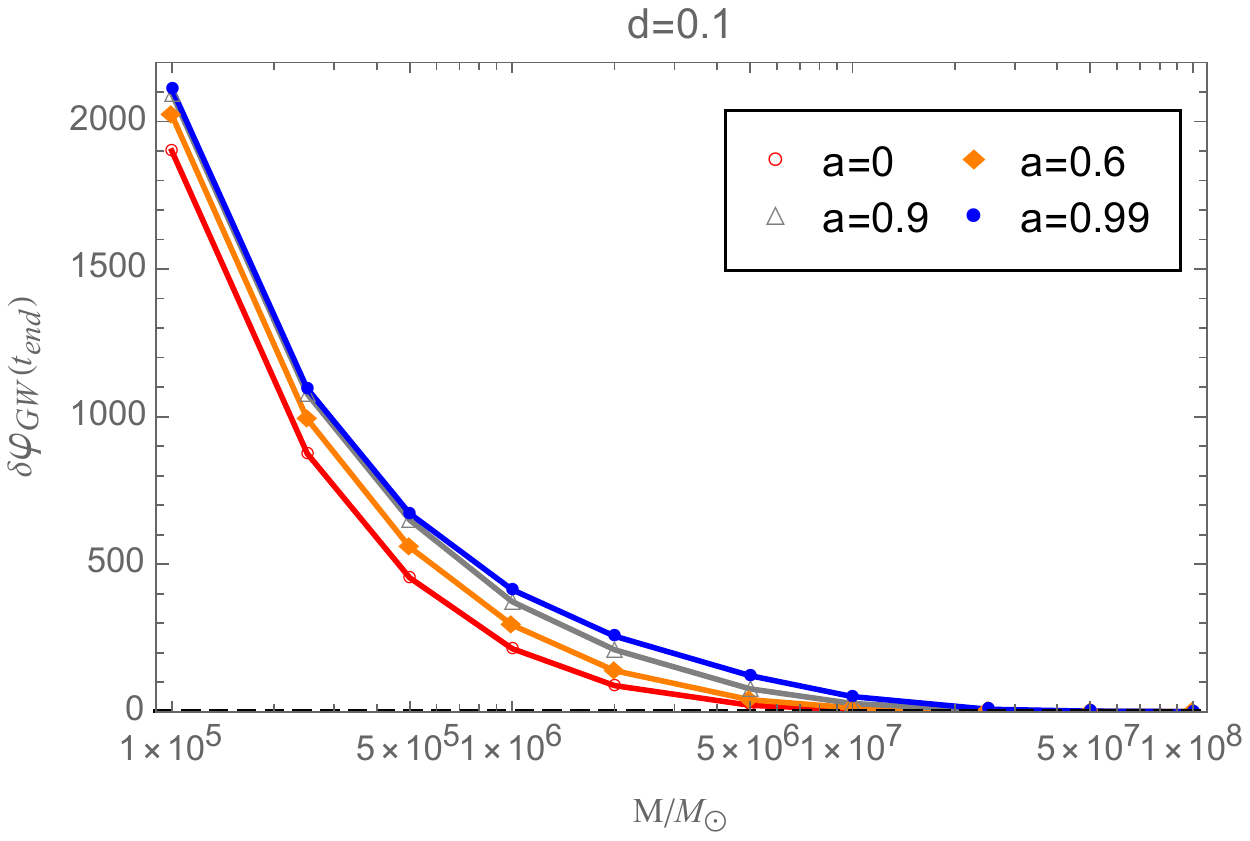}
\includegraphics[scale=0.42]{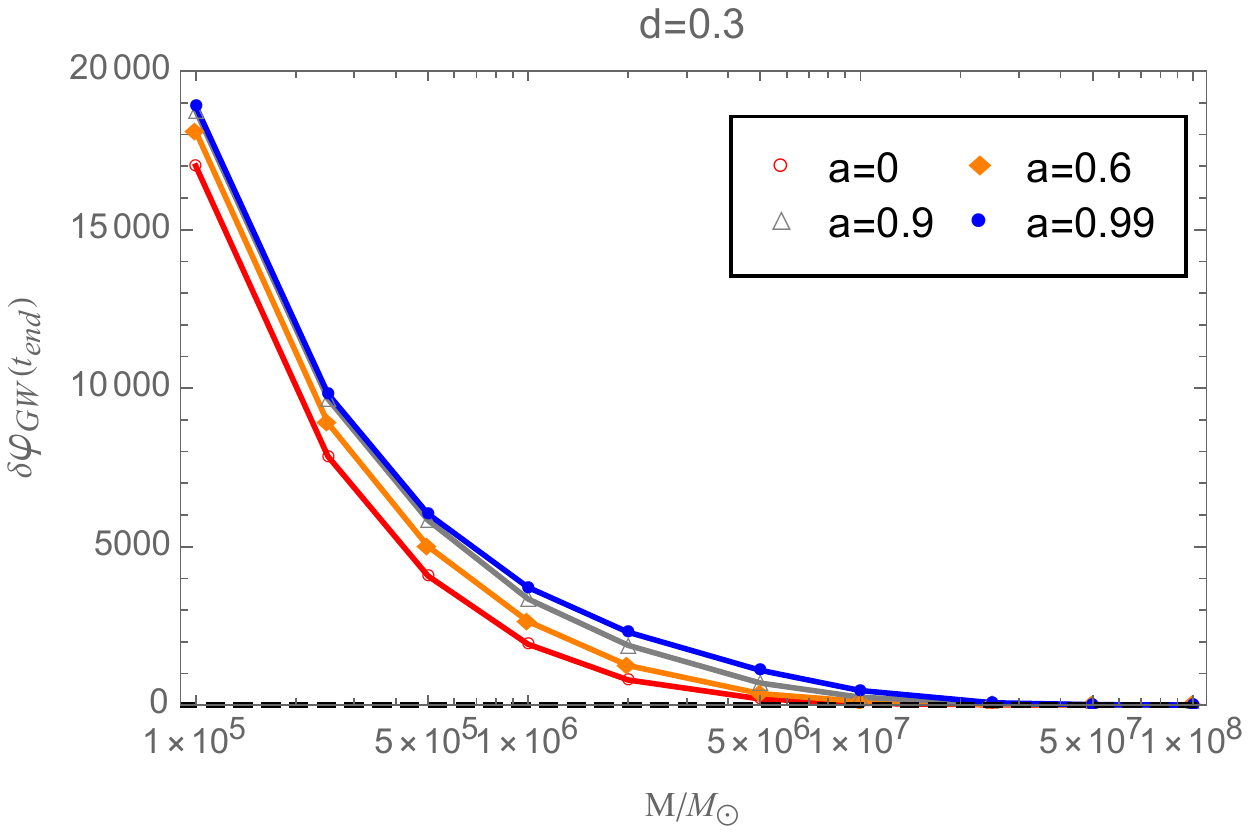}
\caption{Fixing $q$, each panel shows the dependence of dephasing $\delta\varphi_{GW}(t_{\rm end})$ on $M/M_{\odot}$ with different values of $a=0,\ 0.6M,\ 0.9M,\ 0.99M$.
}\label{diffa}}
\end{figure}

\begin{figure}[thbp]
\center{
\includegraphics[scale=0.9]{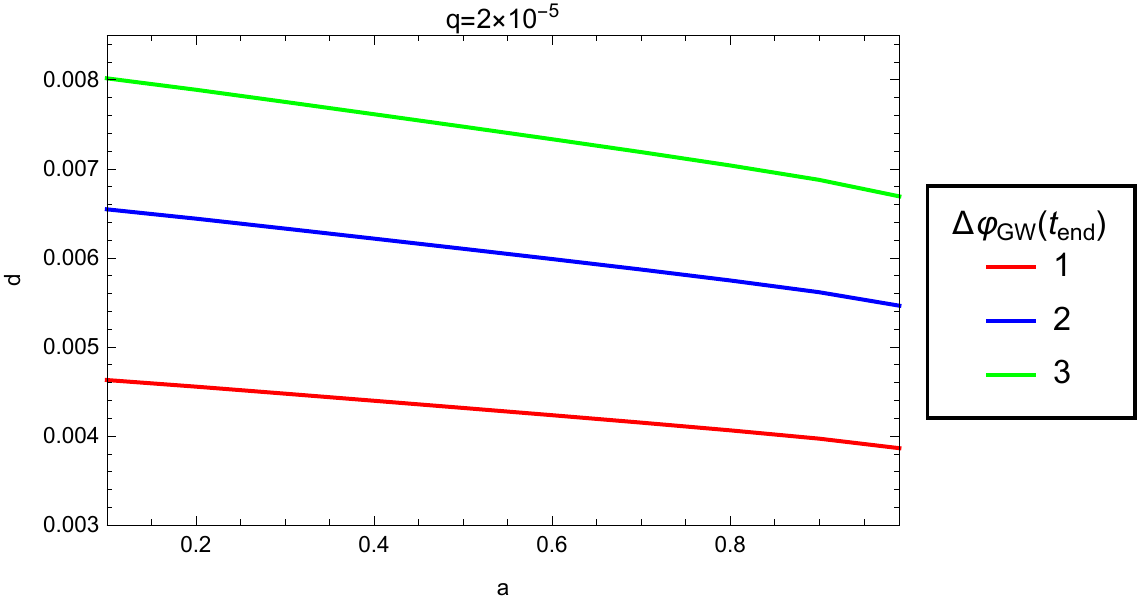}
\caption{Contour plot of gravitational-wave dephasing $\delta\varphi_{GW}(t_{\rm end})$ in the two-dimensional parameter space of $a$ and $d$.
The red, blue, and green contour lines label the isocurve with $\delta\varphi_{GW}(t_{\rm end})=1,2$ and $3$ radians, respectively.
The contour line of one radian is the standard conservative value for a detectable dephasing.
}\label{contour}}
\end{figure}

Figs. \ref{cycles} and \ref{diffa} show the dephasing as a function of mass of supermassive BHs.
The mass of the secondary body is fixed as $m_{\rm p}=10M_{\odot}$.
The black dashed vertical lines in Fig. \ref{cycles} label one radian's position, which is considered as the conservative value for a detectable dephasing.
For each panel, we see that when a BH becomes more massive, 
the dephasing $\delta\varphi_{GW}(t_{\rm end})$ decreases monotonically.
For a chosen mass ratio, Fig. \ref{cycles} shows $\delta\varphi_{GW}(t_{\rm end})$ increases significantly as $d$ increases, while Fig. \ref{diffa} displays that $\delta\varphi_{GW}(t_{\rm end})$ increases with the increase of central BH's spin.
The blue dotted line in the left bottom panel of Fig. \ref{cycles} labels the EMRI with mass ratio $10^{-5}$ and dimensionless spin $a=0.9$, which was discussed in \cite{Maselli:2021men}. With the specific choice of parameters, it was argued that the presence of a scalar charge as small as $d\sim 0.005-0.01$ could be detected in space.
We see that as the primary body loses mass, the dephasing increases for a fixed scalar charge so that the accumulative effect increases as the mass ratio increases.
It is possible to measure a smaller scalar charge in EMRIs with a larger mass ratio.

To show the dependence of dephasing $\delta\varphi_{GW}(t_{\rm end})$ on scalar charge $d$ and central BH's spin $a$, in Fig. \ref{contour} we set $M=5\times 10^5M_{\odot}$, $m_p=10M_{\odot}$,  and plot  $\delta\varphi_{GW}(t_{\rm end})$ in the  parameter space  $a$ and $d$.
The red isoline corresponds to the conservative value for a detectable dephasing, i.e., one radian, while the blue and green curves are the constant value of two and three radians, respectively.
We see that fast rotating BH with larger $a$ can allow to detect smaller  scalar charge $d$, e.g. $d\simeq 0.0048$ for $a=0.1$, $d\simeq 0.004$ for $a=0.99$.
This scenario indicates that dephasing is more sensitive to the scalar charge in EMRIs with a larger $a$.

\begin{figure}[thbp]
\center{
\includegraphics[scale=0.74]{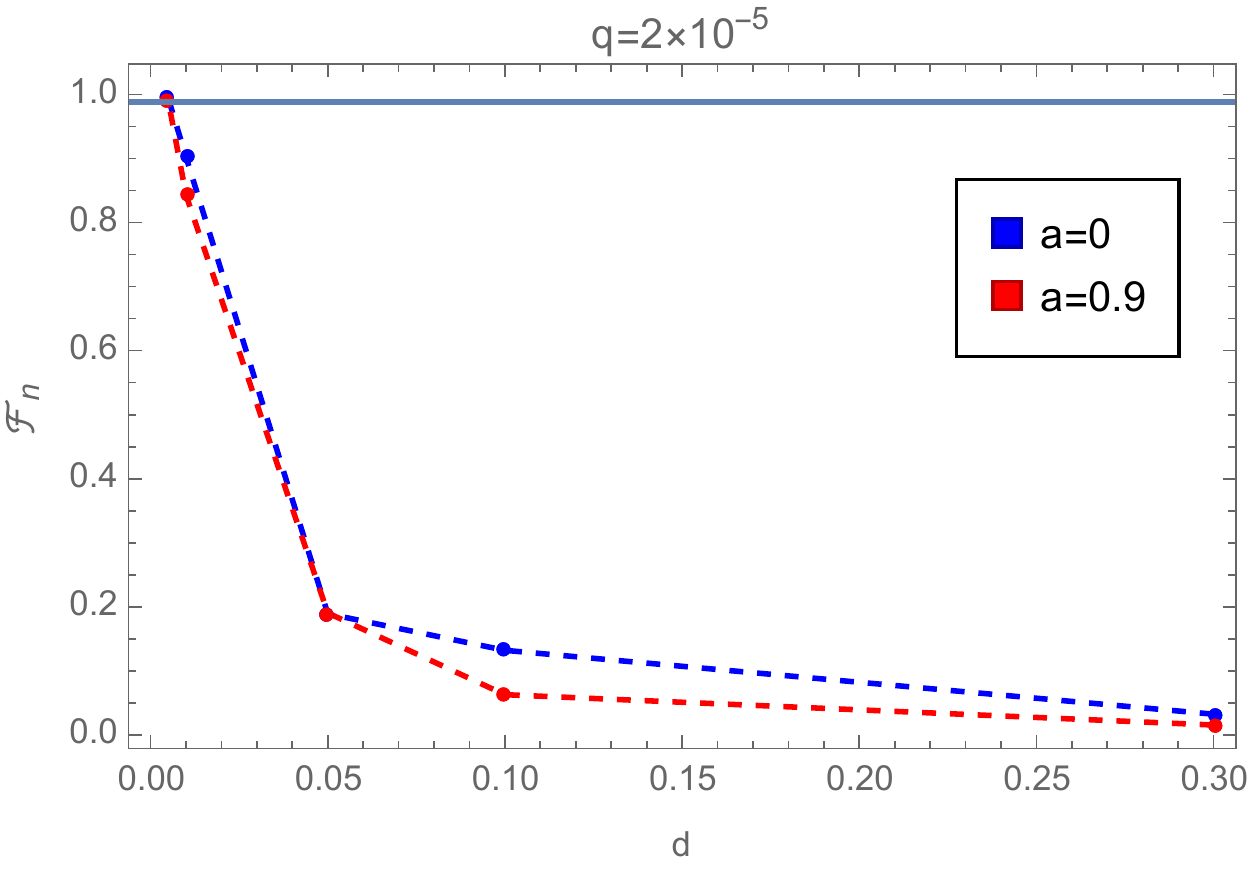}
\caption{Faithfulness between the GW signals with and without of the scalar charge $d$ as the function of the latter.
The calculations are carried out for $a=0$ and $a=0.9$. 
The horizontal solid line represents the detection limit $\mathcal{F}_n\leq 0.988$ by assuming a SNR of $30$.
}\label{faithfulness}}
\end{figure}

By investigating the dephasing, we estimated the effects of the scalar field on EMRIs.
For a more quantitive and accurate assessment, we proceed to analyze the measurability by the future space-borne detector LISA.
A relevant quantity is {\it faithfulness}, first introduced in Ref.~\cite{Maselli:2021men} based on the inner product weighted by the power spectral density of LISA's floor noise~\cite{Lindblom:2008cm}. 
By definition, it provides an estimation of how the two signals are distinct from each other and potentially lead to the detection of the scalar charge.
A comprehensive study has been performed recently in Ref.~\cite{Maselli:2021men}, and the authors demonstrated that LISA could measure up to a few percent of scalar charges.
For the current study, we calculate the faithfulness $\mathcal{F}_n$ between the GW signals in GR and the present model.
To be specific, we consider two values of the spin parameter $a=0$ and $a=0.9M$, while assuming $M=10^6M_{\odot}$ , $q=2\times 10^{-5}$, $\varphi_0=0$, $\theta_S=\phi_S=0.2$, and $\theta_K=\phi_K=0.8$.
For a given SNR $\rho=30$, one can identify the presence of the scalar charge through the GW signals when $\mathcal{F}_n\leq 0.988$. 
We show the results in Fig.~\ref{faithfulness}, where the faithfulness is evaluated as a function of the scalar charge $d$. 
Also, the details regarding the calculations are relegated to the appendix. 
The results presented in Fig.~\ref{faithfulness} largely agree with those found in~\cite{Maselli:2021men}.
It is observed that the calculated faithfulness for $a=0.9M$ always stays below those for $a=0$.
The lower bound for the detection is found to be as small as $d\approx 0.005$. 
These results support the above results on the dephase of GW owing to a nonvanishing scalar charge.

It is also worth mentioning that the above calculations have been mainly focused on the role of the scalar charge.
In other words, the preceding analysis has not taken into account possible degeneracy among different model parameters.
The remaining model parameters may also play a substantial role, leading to a possible degeneracy in the theory's parameter space.
To this end, one may carry out an analysis in terms of the Fisher information matrix~\cite{Vallisneri:2007ev,Gair:2012nm}.
Indeed, this aspect has been convincingly investigated recently in Ref.~[63].
It was shown that a more sophisticated survey of the parameter space does not undermine the feasibility of measuring the scalar charge.
In this regard, as an educated guess, we understand that it is plausible that the above conclusion will continue to be valid under a joint investigation of the entire parameter space.
A more detailed study is left to future work.

Before closing this section, we comment on how the observations might give rise to a more stringent bound on the theories of gravity.
It is well known that the experimental data from the solar system brings relevant constraints on the parameters of the theories of gravity~\cite{Will:2014kxa}, and in particular, in terms of those derived by the parameterized post-Newtonian (PPN) formalism~\cite{Bertotti:2003rm}.
In this context, the Brans-Dicke theories have been largely ruled out by the solar system observations~\cite{Perivolaropoulos:2009ak,Klimek:2009zz}.
On the other hand, the solar system might not be suitable to test the strong, dynamical, and non-linear features of the gravitational interaction~\cite{Yagi:2015oca} owing to the strength of the gravity and magnitude of the planet's velocities.
Therefore, it is not very surprising that calculations carried out for theories with high-curvature corrections, such as Einstein-dilaton Gauss-Bonnet theory~\cite{Sotiriou:2006pq} and 4D Einstein-Gauss-Bonnet~\cite{Clifton:2020xhc} models, have all succeeded in passing the solar test.
In this regard, EMRIs are expected to provide further constraints on the validity of candidate theories of modified gravity, especially for theories featured by strong-curvature modifications.
As a result, joint observations that unite solar observations, future EMRI, and binary system~\cite{Yagi:2015oca} are expected to provide a more stringent constraint on the scalar charge. 
In practice, for a specific modified gravity, the scalar charge $d$ is often related to the parameter of the theory.
By the skeletonization procedure, in Refs.~\cite{Kanti:1995vq,Pani:2009wy,Sotiriou:2013qea,Sotiriou:2014pfa,Julie:2019sab}, for Einstein-dilaton Gauss-Bonnet theory with $f(\phi)\propto e^{\phi}$, it was found that 
\begin{equation}
d=2\beta+\frac{73}{30}\beta^2+\frac{15577}{2520}\beta^3+\mathcal{O}(\beta^4) .\nonumber
\end{equation}
While for shift-symmetric scalar-Gauss-Bonnet theory with $f(\phi)\propto \phi$, one has
\begin{equation} 
d=2\beta+\frac{73}{60}\beta^3+\mathcal{O}(\beta^5) ,\nonumber
\end{equation}
where the dimensionless constant $\beta$ is defined as $\beta=q^{-2}\zeta=\alpha/m^2_p$~\cite{Kanti:1995vq,Pani:2009wy,Julie:2019sab}, where $\alpha$ is the coupling between the curvature and the scalar field and $m_p$ is the mass of the small object.
Subsequently, a measurement of the charge $d$ could imply a bound on the underlying modified gravity.
On the other hand, the model investigated in our study is rather generic and does not necessarily satisfy these relations.
Nonetheless, our simulations indicate for the spin parameter $a\leq 0.99M$, the least detectable scalar charge is approximataly $d\gtrsim 0.004$.
If one further assumes the above relations, it implies that the coupling $\alpha \gtrsim 0.001995 m_p^2$ and $\alpha \gtrsim 0.002 m_p^2$, respectively.

We conclude this section by reiterating that dephasing due to the scalar wave emission has been found to be more significant for EMRIs with either a larger mass ratio or a faster rotating central black hole.
The calculations indicate that EMRI can serve as a sensitive probe to the scalar field in modified gravity.

\section{Further discussions and concluding remarks}\label{sec=conslusion}

We studied the EMRIs in an alternative theory of gravity with an additional scalar field and investigated the imprint of the scalar field on the gravitational waveform.
Besides GWs of tensor modes, scalar radiation also plays a crucial role in the present model. 
The equation of metric perturbation is numerically solved in the Teukolsky formalism, while the equation for scalar perturbation is computed in Green's method.
We simulated the EMRIs up to four years before the merger and explored its dependence on the parameters such as the mass ratio $q$, scalar charge $d$, and central BH's spin $a$.

We computed the total energy flux radiated to the spatial infinity.
It was found that increasing either the scalar charge or the central BH spin enhances the strengths of both gravitational and scalar energy emissions.  
The flux of the GW emission increases more rapidly than that of the scalar wave radiation in the high-frequency region, and the correction $\delta\dot{E}_d/\dot{E}_{GR}$ decreases when a smaller object spirals into ISCO.

The dephasing generated by the scalar field emission was also investigated.
The additional scalar emission increases the total energy flux, resulting in the orbit shrinking faster than that observed in GR.
Compared with the case in GR, a smaller object experiences fewer orbital cycles before the merger for identical initial conditions.
Hence, the orbital phase difference $\delta\varphi_{\rm orb}(t)$ is always positive.
Through a scan in the parameter space, we found that the dephasing changes monotonously with the parameters.
As the mass ratio increases, the dephasing $\delta\varphi_{GW}(t_{\rm end})$ increases. 
The dephasing also increases with the central BH's spin and the secondary object's scalar charge.
Thus we concluded that the dephasing in EMRI due to the scalar radiation is sensitive to the scalar charge when the system possesses a significant mass ratio and fast rotating central BH. 
Since EMRI is more likely to be observable in space-based gravitational observation, the signature of the scalar field can be potentially probed in a BH binary system containing the above characteristics.  

It is interesting to further generalize the present study to the case of generic orbits and examine the effect of the small compact object's spin.
Besides, it is worth exploring more realistic scenarios in astrophysics, inclusively when the central massive BH is surrounded by an accretion disk or immersed in a superlight dark matter cloud.
The specific waveform in such a case is a pertinent topic that is crucial for extracting novel physics from empirical data. 
These open questions call for more studies. 

\appendix

\section{Calculation of the faithfulness}

The GW strain in the space-based detector under the long-wavelength approximation reads
\begin{equation}
\label{signal}
s(t)=\frac{\sqrt{3}}{2}\left[h_+(t)F^+(t)+h_\times(t)F^\times(t)\right],  
\end{equation}
where $h_+(t)=\mathcal{A}\cos\left[2\varphi_{\rm orb}+2\varphi_0\right]\left(1+\cos^2\iota\right)$, $h_\times(t)=-2\mathcal{A}\sin\left[2\varphi_{\rm orb}+2\varphi_0\right]\cos\iota$,  $\iota$ is the inclination angle between the binary orbital angular momentum and the line of sight, and the GW amplitude  $\mathcal{A}=2m_{\rm p}\left[M\Omega(t)\right]^{2/3}/d_L$, $d_L$ is the luminosity distance.
The interferometer pattern functions $F^{+,\times}(t)$ and $\iota$ can be expressed in terms of four angles which specify the source orientation, $(\theta_S,\phi_S)$, and the orbital angular direction $(\theta_K,\phi_K)$ \cite{Katz:2021yft,Cutler:1997ta}.
Further, we take the modulation of Doppler phase due to the LISA orbital motion into  the GW signal \eqref{signal} \cite{Babak:2006uv}
\begin{equation}
\label{signal2}
\varphi_{\rm orb}(t)\rightarrow \varphi_{\rm orb}(t)+\frac{d\varphi_{\rm orb}(t)}{dt}R_{\text{AU}}\sin \theta_S \cos\left(2\pi t/(1~\text{year})-\phi_S\right).
\end{equation}
To estimate the scalar charge distinguishability, one introduces the noise-weighted inner product to define the the faithfulness between two templates
\begin{equation}
\langle s_1|s_2 \rangle=2\int_{f_{\rm min}}^{f_{\rm max}}\frac{\tilde{s_1}(f)\tilde{s_2}^*(f)+\tilde{s_1}^*(f)\tilde{s_2}(f)}{S_{n}(f)}df,
\end{equation} 
where $S_n(f)$ is the noise power spectral density of LISA \cite{Robson:2018ifk} without Galactic binary foreground noise for convenience, $f_{\text{min}}$ and $f_{\text{max}}$  are defined by Eq.(\ref{fmaxmin}).
Faithfulness between two signals can also be determined 
\begin{equation}\label{eq:def_F}
\mathcal{F}_n[s_1,s_2]=\max_{\{t_c,\phi_c\}}\frac{\langle s_1\vert
	s_2\rangle}{\sqrt{\langle s_1\vert s_1\rangle\langle s_2\vert s_2\rangle}}\ ,
\end{equation}
where $(t_c,\phi_c)$ are time and phase offsets \cite{Lindblom:2008cm}.
The two resulting signals can be distinguished by LISA if $\mathcal{F}_n\leq0.988$ \cite{Chatziioannou:2017tdw}.

\begin{acknowledgments}
The authors thank the helpful discussions with Prof. Bin Wang.
This research is supported by the National Key Research and Development Program of China under Grant No.2020YFC2201400.
YG acknowledges the support by the National Natural Science Foundation of China under Grant No. 11875136.
We also acknowledge the financial support from Brazilian agencies 
Funda\c{c}\~ao de Amparo \`a Pesquisa do Estado de S\~ao Paulo (FAPESP),
Funda\c{c}\~ao de Amparo \`a Pesquisa do Estado do Rio de Janeiro (FAPERJ),
Conselho Nacional de Desenvolvimento Cient\'{\i}fico e Tecnol\'ogico (CNPq),
Coordena\c{c}\~ao de Aperfei\c{c}oamento de Pessoal de N\'ivel Superior (CAPES).
\end{acknowledgments}

\bibliographystyle{JHEP}
\bibliography{EMRItest}
\end{document}